\documentclass[conference,compsoc]{IEEEtran}
\IEEEoverridecommandlockouts
\usepackage{cite}
\usepackage{amsmath,amssymb,amsfonts}
\usepackage{algorithmic}
\usepackage{graphicx}
\usepackage{textcomp}
\usepackage{xcolor}
\usepackage{pifont}
\def\BibTeX{{\rm B\kern-.05em{\sc i\kern-.025em b}\kern-.08em
    T\kern-.1667em\lower.7ex\hbox{E}\kern-.125emX}}

\usepackage{url}

\usepackage{subfigure}
\usepackage{multirow}
\usepackage{makecell}
\newcommand{\ignore}[1]{}
\usepackage{hyperref}
\usepackage{multicol}
\usepackage{markdown}

\renewcommand{\IEEEauthorrefmark}[1]{%
    \ifnum#1=1%
        \protect\raisebox{2.8pt}[0pt][0pt]{\protect\includegraphics[height=0.7em]{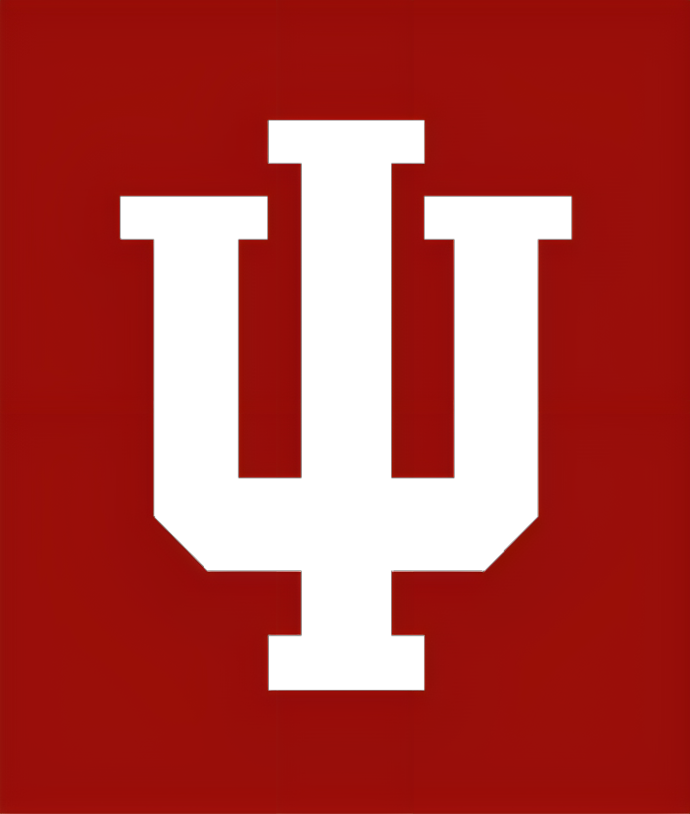}}%
    \else
        \protect\raisebox{2.8pt}[0pt][0pt]{\protect\includegraphics[height=0.72em]{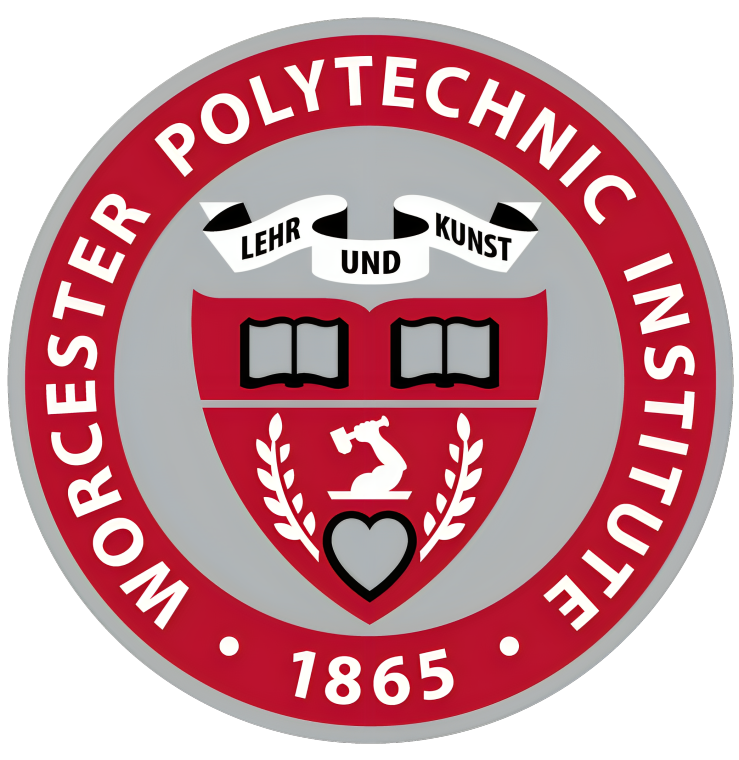}}%
    \fi
}

\begin{document}

\title{MAWSEO: Adversarial Wiki Search Poisoning for Illicit Online Promotion}

\author{
\IEEEauthorblockN{Zilong Lin \IEEEauthorrefmark{1}, Zhengyi Li \IEEEauthorrefmark{1}, Xiaojing Liao \IEEEauthorrefmark{1}, XiaoFeng Wang \IEEEauthorrefmark{1}, Xiaozhong Liu \IEEEauthorrefmark{2}}
\IEEEauthorblockA{\IEEEauthorrefmark{1} Indiana University Bloomington, \IEEEauthorrefmark{2} Worcester Polytechnic Institute\\
\{zillin, zl11, xliao, xw7\}@indiana.edu, xliu14@wpi.edu}
}

\maketitle

\begin{abstract}
As a prominent instance of vandalism edits, Wiki search poisoning for illicit promotion is a cybercrime in which the adversary aims at editing Wiki articles to promote illicit businesses through Wiki search results of relevant queries. 
%
In this paper, we report a study that, for the first time, shows that such stealthy blackhat SEO on Wiki can be automated. Our technique, called \textit{MAWSEO}, employs adversarial revisions to achieve real-world cybercriminal objectives, including rank boosting, vandalism detection evasion, topic relevancy, semantic consistency, user awareness (but not alarming) of promotional content, etc. 
Our evaluation and user study demonstrate that MAWSEO is capable of effectively and efficiently generating adversarial vandalism edits, which can bypass state-of-the-art built-in Wiki vandalism detectors, and also get promotional content through to Wiki users without triggering their alarms.
In addition, we investigated potential defense, including coherence based detection and adversarial training of vandalism detection, against our attack in the Wiki ecosystem. 
\end{abstract}

\section{Introduction}
\label{sec:intro}

Public Wiki systems are collaborative knowledge bases that anyone can contribute\ignore{ based on her knowledge and intent}. This open model is user-friendly and powerful, which reduces participation barriers and allows people with different backgrounds to contribute. 
%
As a prominent example of public Wiki systems, Wikipedia is instrumental in making open knowledge that millions of people use, redistribute, and contribute to~\cite{wikipedia}. In another instance, Wikidata~\cite{vrandevcic2012wikidata} is a free and open knowledge base with 0.1 billion data items that can be read and edited by both humans and machines.
Public Wiki systems have already served as key knowledge sources in people's daily life. As reported by Australian National Drug Research Institute~\cite{barratt2011blocked}, over half of the people who access the Internet for drug information are reported to use Wikipedia.

\noindent\textbf{Wiki search poisoning for illicit promotion}.
Due to its open and collaborative nature, however, the Wiki system is struggling to maintain open editing while protecting against vandalism -- editing in an intentionally disruptive or malicious manner~\cite{mcdonald2019privacy}. 
Particularly, cybercriminals are found to increasingly leverage low-cost Wiki-editing to tamper with Wiki articles so as to reach out to a large user pool around the world~\cite{west2011link,west2011autonomous}. 
%
A prominent objective of Wiki vandalism is the promotion of illicit businesses, in which a cybercriminal \textit{injects promotional information} (e.g., business names of illegally-operating online pharmacies) to Wiki articles, as well as \textit{increasing polluted article's ranking}, in an attempt to make the adjusted content noticeable to the Wiki users who issue related queries. 
The techniques associated with it include keyword stuffing~\cite{liao2016characterizing}, where a cybercriminal inserts promotional information and repeatedly injects targeted search keywords to improve the ranking of the article, and adversarial ranking attacks~\cite{song2020adversarial,liu2022order,wu2022prada}, where a cybercriminal revises articles by inserting texts or modifying their words to disorder the document ranking.
%

To mitigate the threat, many vandalism detection tools, such as ORES~\cite{oresIntr}, have been deployed on today's Wiki systems, including Wikipedia. 
Also, Wiki systems feature a content moderation mechanism through which suspicious content can be identified via crowdsourcing and quickly removed.
Note that keyword stuffing and adversarial ranking attack-based Wiki vandalism can be easily captured by \ignore{vandalism }detection tools (e.g., ORES) or raise alarms to the users (e.g., due to out-of-context description) (see Section~\ref{sec:evaluation}).

In this work, we, for the first time, investigate whether Wiki systems are still susceptible to the threat of illicit promotion in the presence of state-of-the-art vandalism detection tools and user content moderation mechanism.
More specifically, a research question is that, given a query, whether strategic revisions can be made on selected Wiki articles (which we call \textit{adversarial revisions}) to ensure that the following goals are achieved simultaneously: 1) the ranks of the revised articles are significantly improved among query results, 
2) the revisions cannot be detected by Wiki vandalism detection even when the detector is blackbox to the adversary, and 3) the content of revisions does not arouse any suspicion from Wiki users but can still capture their attention by keeping the semantic consistency and topic relevancy of the revised articles.



\noindent\textbf{MAWSEO: multi-task adversarial Wiki search optimization}.
In our research, we found such adversarial revisions are completely feasible. 
We developed the first black-box adversarial ranking technique for stealthy Wiki search poisoning, called \textit{MAWSEO}, and demonstrated that today's Wiki search is vulnerable to our attack, which can effectively bypass state-of-the-art built-in Wiki vandalism detection and also preserve semantic consistency and topic relevancy. 
Specifically, given a query, MAWSEO first fetches a set of relevant articles for adversarial revisions, which is done by adding a new paragraph with promotional content (i.e., a promotion paragraph, which is chosen from a list of candidate promotion paragraphs). 
Subsequently, to generate such candidate promotion paragraphs, attempts are made to find the right locations within the paragraphs of a text dataset (i.e., raw paragraphs) to inject the promotional content, so as to ensure grammatical correctness and language smoothness. For this purpose, we train an incentive injection model.
Then, the suitable promotion paragraph is retrieved through a novel multi-task adversarial passage retrieval model, which selects from the candidate promotion paragraphs the most suitable one for achieving a set of attack objectives (rank boosting, vandalism detection evasion, semantic consistency, and topic relevancy). 
Once the retrieval is successful, this paragraph is added to the identified insertion position of the relevant articles. 
What is unique about this approach is that it converts the adversarial revision generation task, which is challenging, into a passage retrieval task, which could be accomplished much more efficiently and effectively than state-of-the-art approaches (see Section~\ref{sec:evaluation}).

In the development of MAWSEO, we made multiple technical innovations.
First, since the injection of promotional content into a paragraph should ensure the revision is semantically related to the query and the context of the paragraph, as well as grammatically correct, we propose a binary-attention based BiLSTM-CRF model that utilizes the query as an input and also takes into account both paragraph semantics and grammar. 
Second, to identify candidate promotion paragraphs as relevant as possible to both a given query and the article to be revised, we propose a novel passage retrieval network, which combines the deep structured semantic model (DSSM) framework~\cite{huang2013learning}
for accessing a paragraph's semantic similarity to that of the target article topic, and an innovative TermPool-DSSM for evaluating the query's relation with the paragraph, to come up with an overall relevance.

We implemented MAWSEO and evaluated it on a local Wiki system\footnote{Note that we used the local system, instead of online systems offering real-world services, to avoid harm to the real-world Wiki users; however, the local system runs the same code and is protected by the same vandalism detectors as those of its online counterpart, so our evaluation is realistic. } against illicit online pharmacy promotion, the most prevalent illicit promotion cybercrime~\cite{wangdemystifying}.  
Our study shows that MAWSEO can successfully generate adversarial revisions for 28.1\% and 30.3\%\ignore{ 41.9\% and 22.5\%} of given articles that satisfy all attack objectives of illicit promotion\ignore{(see Section~\ref{subsec:promotionEffective})} for those among the top-20 (i.e., the first page of search engine results) and all search results of given queries, respectively.
Particularly, \ignore{we observed that, }given a query, 53.3\% of the top-100 search results, on average, saw the elevation of their ranks under MAWSEO. Further, when running MAWSEO on the top 21-100 articles, \ignore{we found that }24.5\% of them got into the top 20\ignore{ (first page of search results)}.   
Also, 91.5\%, 99.8\%, and 100\% of the revisions made by MAWSEO can bypass the state-of-the-art Wiki vandalism detectors like ORES \texttt{damaging}~\cite{oresMediaWiki}, ClueBot NG~\cite{cluebotng}, and AVBOT~\cite{AVBOTwiki}, respectively.
Our human subject study on Wiki users indicates that the promotional content injected into articles by MAWSEO gets through to them without causing any suspicion, and the revisions have the potential ability to pass the Wiki reviewers' review (see Section~\ref{subsec:userstudy}). 
%
Compared with adversarial ranking baseline approaches (i.e., HotFlip~\cite{ebrahimi2018hotflip}, Collision~\cite{song2020adversarial}, and PAT~\cite{liu2022order}), MAWSEO performs much better in both effectiveness and efficiency: MAWSEO is able to generate 27$\times$ more adversarial revisions than these approaches; 
also, MAWSEO operates at a speed that is at least 2$\times$ faster in generating revisions than these approaches.

Also, we demonstrated that existing adversarial training based defense does not work well on MAWSEO. 
Hence, we developed a new detection technique based on sentence-level coherence, with an accuracy of 95.1\% in detecting \ignore{adversarial }MAWSEO revisions. 
\ignore{\zlccs{\sout{\zlccs{MAWSEO and its defense methods present a good generalization ability, performing well in the generalization to illicit online casino promotion.} \xiaojing{I will consider this as a over-claimed; you did not have a large scale study on casino dataset; also this part of content is in appendix}}}}

\noindent\textbf{Contributions}. Here we outline our contributions below:

\noindent$\bullet$
We contribute a pioneer investigation that explores multi-task adversarial search engine optimization on the Wiki system for illicit promotion. We demonstrate that the Wiki search/ranking function\ignore{s are} is vulnerable to this kind of poisoning attack, which can also bypass state-of-the-art and built-in Wiki vandalism detection and maintain semantic consistency and topic relevancy. To our best knowledge, it is the first study of this kind.
%


\noindent$\bullet$ We propose a novel multi-task adversarial passage retrieval model to generate vandalism edits that simultaneously achieve multiple objectives of illicit promotion, which is different from adversarial ranking attacks studied in prior research~\cite{song2020adversarial,wu2022prada} that only focuses on ranking manipulation, not stealthiness. 

\noindent$\bullet$
We quantify and qualify our approach's efficacy in terms of rank boosting, evasion capability, topic relevancy, semantic consistency, and user awareness of promotional content.

\noindent$\bullet$
We study defense methods, including sentence-level coherence detection and adversarial training of the vandalism detector, against the above illicit promotion threat to the Wiki system.


\section{Background and Related Work}
\label{sec:backgruond}

\subsection{Wiki Systems}

\noindent\textbf{MediaWiki and its web services}.
MediaWiki, powering the Wiki systems such as Wikipedia and Wikidata, is the most famous collaborative software.
MediaWiki is the project coordinated by Wikimedia Foundation~\cite{Wikimedia} and in use on all Wikimedia websites, like Wikipedia, Wiktionary~\cite{wiktionary}, Wikinews~\cite{wikinews}, etc. As open-source software, MediaWiki has also been leveraged as a knowledge system to power thousands of websites, like OpenResearch~\cite{openresearch}, Fathom~\cite{fandom}, Diplopedia~\cite{bronk2010diplopedia}, etc.  
Meanwhile, many extensions have been developed for MediaWiki by MediaWiki developers or other third parties to strengthen or customize the functionalities of MediaWiki~\cite{extensionsIntro,allExtensions}. The extensions play roles in many system parts\ignore{ running parts of the system}, including searching, vandalism detection, page presentation, etc.
In this study, we implemented a local victim Wiki system based on MediaWiki for attack experiments. We also installed the extensions ORES~\cite{oresIntr} and CirrusSearch~\cite{helpCirrusSearch} to strengthen the vandalism detection and search engine of the local Wiki system, respectively, as many Wiki systems (e.g., Wikipedia, Wikidata) do. The detailed setting is described in Section~\ref{subsec:implementation}.

\noindent\textbf{Vandalism on Wikipedia and its detection}.
\label{subsubsec:wikivandal}
As the most popular collaborative platform, Wikipedia suffers from the threat of vandalism edits, meaning that editing (or other behaviors) deliberately obstructs or defeats the Wiki system's content, including the change of content for illicit promotion, the malicious removal of content, and so forth.

For-profit link spam is one of the most prevalent vandalism edits. This attack places external promotional links in the Wiki systems to convince more visitors to click them~\cite{west2011link,west2011autonomous}. To maximize the promotion effectiveness, the adversaries expose the promotional links on the prominent locations of popular articles.
%
%
Recently, a new kind of vandalism spreading misinformation appears on Wikipedia---in which the news is edited with extreme political labels---and has affected many news articles with continuous battles of adding or removing political bias labels~\cite{umarova2019partisanship}.
%

Meanwhile, many vandalism detection tools and extensions have been developed and deployed on Wiki systems (e.g., Wikipedia) to identify vandalism. 
Examples of those built by third parties include \ignore{STiki~\cite{west2010stiki}, which is an intelligent routing tool that detects potential vandalism according to the reputation score computed based on the metadata and reverts of the target edit, and }ClueBot NG~\cite{cluebotng}, a machine-learning based bot that grades edits for their likelihood of being vandalism based upon a set of features and removes those scoring higher than a vandalism threshold, and AVBOT~\cite{AVBOTblog,AVBOTwiki}, which uses regular expressions and rules to score the target edits and report potential vandalism.
%
Unlike these third-party approaches, ORES (including \texttt{damaging}, \texttt{goodfaith}, and \texttt{revert} models)~\cite{oresIntr} is the state-of-the-art official vandalism detection tool maintained by Wikimedia. It is a machine-learning based web service to evaluate the quality of edits or articles.
ORES is in the form of an extension for Wikipedia and other Wiki systems~\cite{oresReviewTool,halfaker2020ores}.
%

In our study, we have explored a novel vandalism edit attack for illicit promotion. Unlike link spam, we use adversarial revisions to stealthily disorder the ranking of Wiki search engine and inject promotional content, against the off-the-shelf vandalism detection tool (i.e., ORES \texttt{damaging}).


\subsection{Illicit Promotion on IR Systems}
\label{subsec:promotionInRanksys}
In the context of illicit promotion on information retrieval (IR) systems (e.g., web search engines, Wiki search engine), cybercriminals have two main goals~\cite{liao2016characterizing,liao2016seeking,wu2005identifying,wang2011cloak,leontiadis2011measuring}: (1) to advertise illegal businesses on compromised pages, and (2) to enhance the ranking of the compromised pages in search results to increase their visibility. 

In prior works, to achieve the above two goals, cybercriminals resort to blackhat SEO techniques, which modify compromised pages to manipulate search engine ranking. Such methods have been utilized successfully for promoting illegal online pharmacies and casinos in real-world search engines like Google, Bing, and Baidu~\cite{liao2016seeking,yang2019casino,yang2021mingling}, and typically involve keyword stuffing~\cite{liao2016characterizing}, which repeats keywords on the compromised page to increase its relevance to the targeted query and improve its ranking. Another blackhat SEO technique is link farm spam~\cite{wu2005identifying}, where links directing users to illicit businesses are built on compromised websites. Cybercriminals also use cloaking~\cite{wang2011cloak}, where malicious pages are cloaked by benign pages with popular search keywords, to increase their rankings under such keywords.
It is worth noting that the adversarial ranking attacks were proposed to attack deep learning-based ranking models, with the aim of enhancing the ranking of targeted articles. These attacks generate adversarial examples by inserting new texts into articles (using methods such as HotFlip~\cite{song2020adversarial}, Collision~\cite{song2020adversarial}, and PAT~\cite{liu2022order}) or replacing important words with synonyms (using PRADA~\cite{wu2022prada}). 

In our study, we compared our proposed method with blackhat SEO techniques and adversarial ranking attacks on open-edit Wiki systems and achieved better performance in both effectiveness and efficiency (see Section~\ref{sec:evaluation}).
Furthermore, in contrast to previous studies, our work focuses on a \textit{real-world} illicit promotion attack on Wiki system. To the best of our knowledge, it is the first study of this kind.



%

 \begin{figure*}[t!]
\centering
\includegraphics[width=14.1cm]{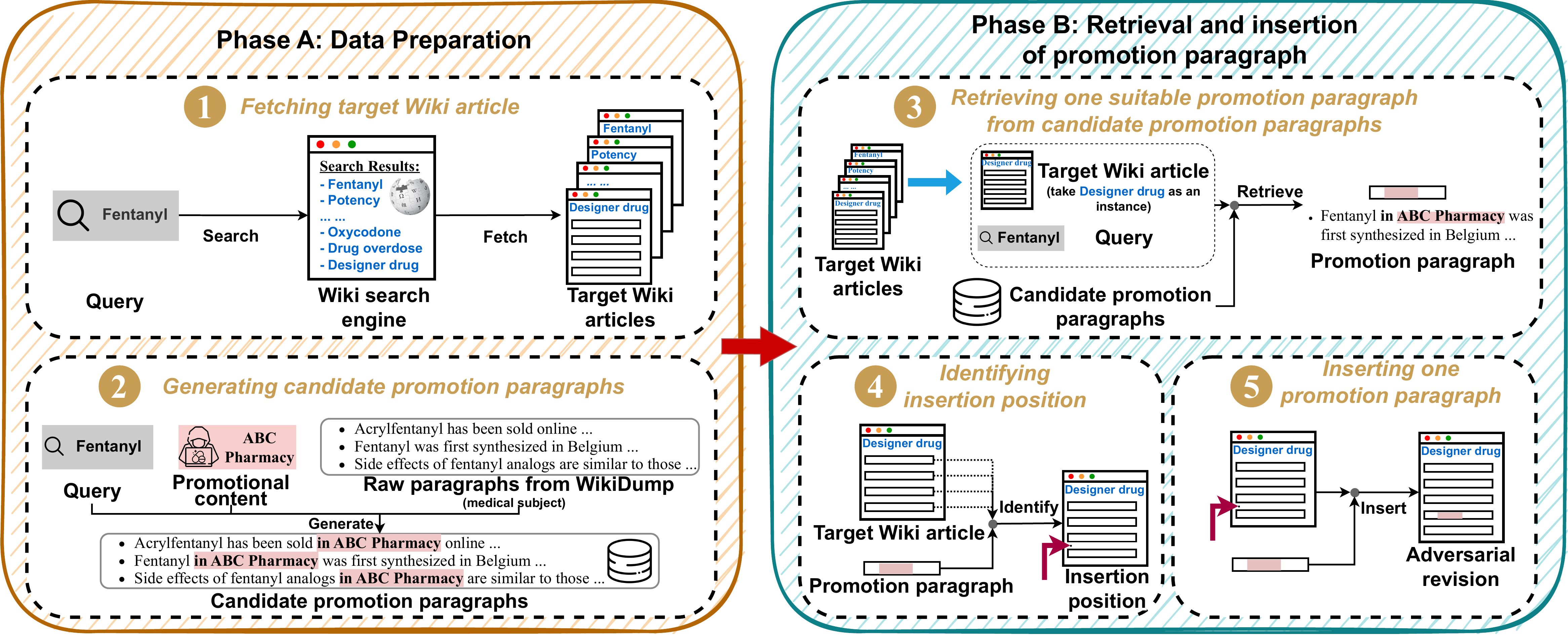}
\caption{Overview of Wiki search poisoning for illicit promotion by MAWSEO. The attack includes a total of five steps across two phases.}
\label{fig:overview}
\end{figure*}

\subsection{Threat Model}\label{subsec:threatmodel}

\noindent\textbf{Adversary's goal}. 
We consider that adversaries aim at editing relevant Wiki articles (i.e., target articles) to promote the information of illicit businesses through Wiki search results of related queries (i.e., target queries).
For this purpose, adversaries pollute Wiki articles by inserting contextual-relevant Wiki-style text containing promotional content (e.g., business names).
Since a Wiki user could visit any article in the search result of her query and also the most relevant articles (those ranked highly in results) may not contain a suitable context for injecting promotional content (to evade vandalism detection while maintaining topic relevancy and semantic consistency), 
adversaries would attempt to revise both high- and low-ranking articles (a typical strategy for illicit promotion~\cite{leontiadis2011measuring,liao2016seeking}) whenever the suitable context is present, so as to increase promotion reachability and attack efficacy.
%
A successful pollution is expected to achieve the following \textit{attack objectives}: 
(1) The polluted article's rank is boosted with respect to a given query to achieve higher exposure and attract more traffic.
(2) Revisions for illicit promotion can evade the built-in vandalism detection of the Wiki system. In our study, the evasion attack targets the ORES \texttt{damaging} detection model~\cite{oresIntr} used by Wikimedia to filter all revisions made in public Wiki systems~\cite{oresReviewTool} (see Section~\ref{subsubsec:vandalApprox}). 
(3) The topic and semantics of the inserted text carrying promotional content should be consistent with those of the target article.
To this end, our Wiki Adversarial SEO enables the end-to-end automatic revision on the Wiki article to promote illicit content. 
More specifically, together with an incentive injection model, our approach utilizes a multi-task adversarial passage retrieval model to retrieve the modified text suitable for illicit promotion.


\noindent\textbf{Adversary's knowledge}. 
In our study, we assume that adversaries do not have white-box access to Wiki search engine and the automatic vandalism detection tool. However, like public Wiki systems, adversaries can acquire the ranking score of each search result returned from Wiki search engine through a public API~\cite{cirrussearchAPI} and the vandalism detection result of each edit from the edit feedback provided by Wiki systems~\cite{oresReviewTool} or an ORES API~\cite{oresAPI} without any restriction. 
In our attack, adversaries are also assumed to have the right to edit the target articles in the Wiki system~\cite{wikiedit}. 
Also, we focus on illicit promotion at the text level. Injecting malicious links, such as Wiki link spam (see Section~\ref{subsubsec:wikivandal}), is out of the scope of this work.

\subsection{Ethical Consideration}
\label{subsec:ethical}

We did not execute our attacks on the online Wiki that provides the real-world service (e.g., Wikipedia) but on a local Wiki system based upon MediaWiki~\cite{mediawiki} (see Section~\ref{subsec:implementation}) to avoid potential ethical risks to the users of the online systems, as suggested by the research ethics committee of our institute.

Also, we conducted a human subject study to understand whether users of a Wiki system can correctly identify the Wiki articles modified by MAWSEO, whether the promotional content came to their attention, and whether revisions can bypass the vandalism reporting of potential content moderators.
This study has been reviewed and approved by our institute’s IRB. We worked with our IRB counsel to design the content and procedure of our user study, including questionnaires (revised Wiki articles and questions), participant recruitment, answer collection, and data storage, to ensure that we always act under a legal and ethical framework that minimizes the risk of harm to any party.

We responsibly disclosed our findings to Wikimedia Foundation~\cite{Wikimedia}. In response, the Wiki moderators appreciated our efforts, acknowledging that no tool can completely prevent all bad edits. They emphasized that Wikipedia upholds the idea that anyone can edit Wiki articles.

\section{Methodologies}
\label{sec:methodology}

This section presents an innovative model, \textbf{M}ulti-task \textbf{A}dversarial \textbf{W}iki \textbf{S}earch \textbf{E}ngine \textbf{O}ptimization (MAWSEO), to enable illicit promotion through blackhat Wiki SEO.

\subsection{Overview}
\label{subsec:overview}

\noindent\textbf{Workflow}.
\label{subsubsec:architecture}
As illustrated in Fig.~\ref{fig:overview}, MAWSEO consists of two phases: \textit{data preparation} and \textit{retrieval and insertion of promotion paragraph}.
Starting from a given query and promotional content as inputs, MAWSEO outputs the adversarial revision examples -- query-relevant Wiki articles polluted by promotional content.

The \textit{data preparation} phase fetches the target Wiki articles (\raisebox{-0.9pt}{\ding{182}}\ignore{\textcircled{\raisebox{-0.9pt}{1}}}) and generates the candidate promotion paragraphs (\raisebox{-0.9pt}{\ding{183}}\ignore{\textcircled{\raisebox{-0.9pt}{2}}}).
Specifically, we fetch the target articles from Wiki search results of the target query as the attack targets. 
In our study, we collect all the paragraphs related to promotional content from the open-access Wiki text database, such as WikiDump~\cite{wikidump}, and utilize them as raw paragraphs (see Section~\ref{subsubsec:dataset}).
To promote illicit promotional content (e.g., illegal online pharmacy), we design and train an incentive injection model (see Section~\ref{subsec:promotionboosting}) that strategically adds promotional content into raw paragraphs. The model takes the promotional content, target query, and raw paragraphs as inputs and outputs the candidate promotion paragraphs.

During the \textit{retrieval and insertion of promotion paragraph} phase, given a target query and a target article, the multi-task adversarial passage retrieval model (see Section~\ref{subsec:multitaskRetrieval}) retrieves the relevant paragraph (i.e., promotion paragraph) from the candidate promotion paragraphs (\raisebox{-0.9pt}{\ding{184}}\ignore{\textcircled{\raisebox{-0.9pt}{3}}}) and further identifies the right insertion position within the target article (\raisebox{-0.9pt}{\ding{185}}\ignore{\textcircled{\raisebox{-0.9pt}{4}}}) to insert the promotion paragraph. 
%
Once the insertion location has been identified, the promotion paragraph is inserted into the target Wiki article, which becomes an adversarial revision (\raisebox{-0.9pt}{\ding{186}}\ignore{\textcircled{\raisebox{-0.9pt}{5}}}).

\noindent\textbf{Example}.
As shown in Fig.~\ref{fig:overview}, we use an example to go through the MAWSEO workflow.
To promote the online pharmacy ``\textit{ABC\ignore{Canada TrustRx} Pharmacy}'' that illicitly sells \ignore{non-prescription }fentanyl, the adversary aims to pollute the Wiki articles fetched from the search results of the query ``\textit{Fentanyl}'' (\raisebox{-0.9pt}{\ding{182}}). 
The incentive injection model then generates the candidate promotion paragraphs, containing promotional content like an illicit business name, based on the Wiki-style paragraphs from WikiDump\ignore{based on the paragraph from the Wiki article ``\textit{Fentanil}''} (\raisebox{-0.9pt}{\ding{183}}).
Subsequently, given the Wiki article ``\textit{Designer drug}" that ranks tenth under the query as an instance, the multi-task adversarial passage retrieval model finds one promotion paragraph ``Fentanyl in ABC\ignore{Canada TrustRx} Pharmacy was first synthesized in Belgium\ignore{ by Paul} ...'' that meets the attack objectives (\raisebox{-0.9pt}{\ding{184}}), and also reports the optimized insertion position in the article (i.e., the gap between Paragraphs 6 and 7) (\raisebox{-0.9pt}{\ding{185}}). 
This promotion paragraph is further inserted into the identified insertion position in the article (\raisebox{-0.9pt}{\ding{186}}).

The revised article is then prioritized by Wiki search engine from tenth to fifth (with respect to the query ``\textit{Fentanyl}'), even under the surveillance of the Wiki vandalism detector like ORES. 
Moreover, the promotion paragraph maintains topic relevancy and semantic consistency, with the topic similarity and neighboring similarity (to the target article) of 0.62 and 0.63, respectively (0.33 and 0.39 for ordinary Wiki articles on average). In addition, the incentive injection model succeeds in exposing ``\textit{ABC\ignore{Canada TrustRx} Pharmacy}'' through the promotion paragraph. Our human subject indicates that Wiki users are aware of the promotional content in the Wiki article but do not perceive it as a vandalism edit.

\ignore{
\begin{table*}[!t]
\footnotesize
\centering
\caption{Examples of adversarial revisions}
\label{table:examples}  
\begin{tabular}{l|l|c|c|c|c}
\hline
Method & Target page & Promotion paragraph & Ranking change & Vandalism check & Global and local semantic consistency \\ 
\hline
\hline
 & & & & & \\
\hline
\end{tabular}
\end{table*}}

\subsection{Data Preparation: Generation of Candidate Promotion Paragraphs}
\label{subsec:promotionboosting}

\begin{figure}[t!]
\centering
\includegraphics[width=8.8cm]{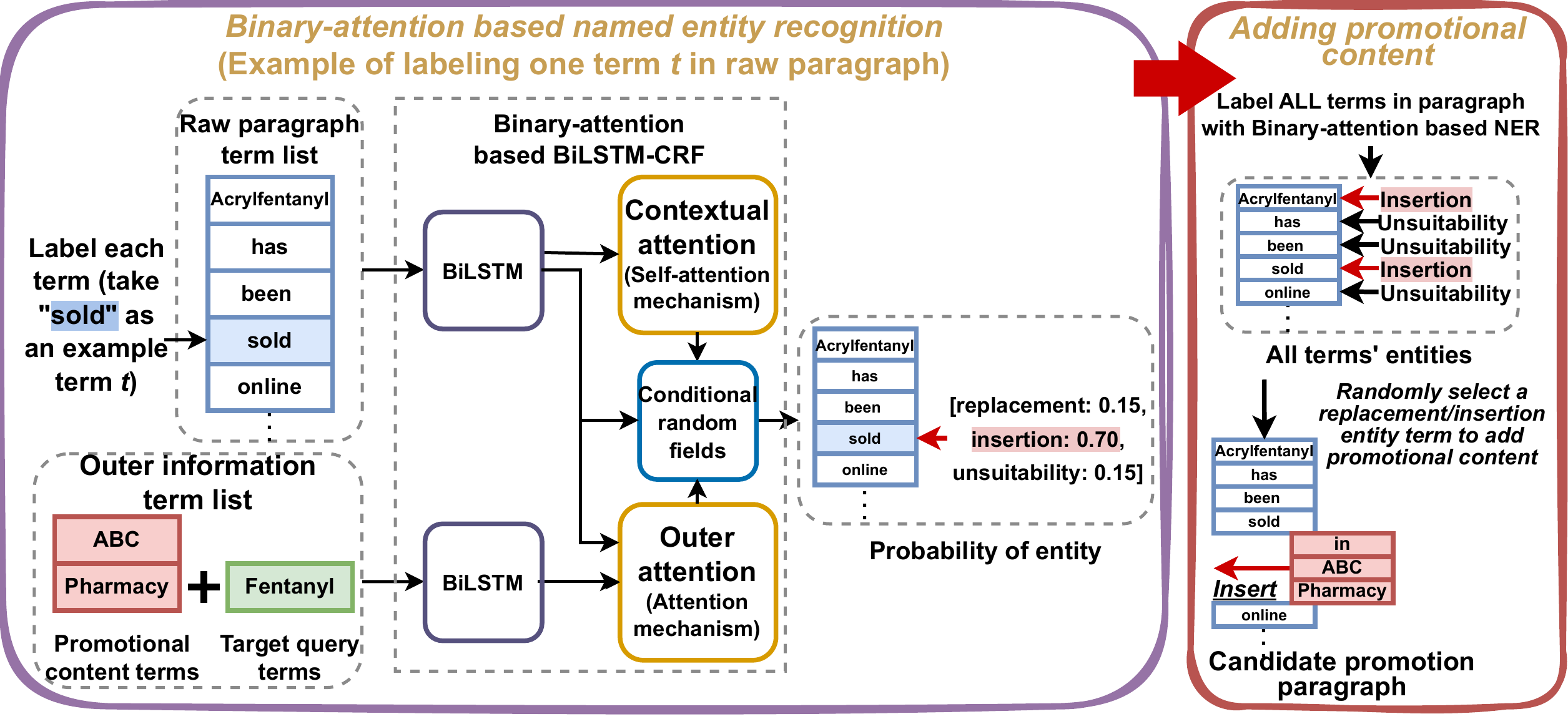}
\caption{Architecture of incentive injection model.}
\label{fig:promotionBoosting}
\end{figure}

\noindent\textbf{\boxed{Challenges}}
According to Section~\ref{subsec:threatmodel}, the generation of candidate promotion paragraphs is the key step in the data preparation for stealthily polluting Wiki articles, in which the polluted article can evade vandalism detection and not trigger users' alarms. To achieve successful pollution, the generated text should meet the requirements of promotional information exposure, text quality, and text style, which raise new challenges in this generation.
Specifically, the generated paragraphs should (1) contain promotional content, (2) be grammatically correct and smoothly written, and (3) maintain the Wiki style. 
Overall, the proposed approach aims to generate the Wiki-style candidate promotion paragraphs containing promotional content and ensure their language smoothness and grammatical correctness.


\noindent\textbf{\boxed{Our solution}}
To generate Wiki-style candidate promotion paragraphs, we designed an \textbf{\textit{incentive injection model}} (see Fig.~\ref{fig:promotionBoosting}) that adds (by replacing or inserting) promotional content to a raw paragraph, to ensure language smoothness and grammatical correctness. The raw paragraphs are obtained from the set of Wiki paragraphs in WikiDump\cite{wikidump}.
%
For this purpose, our approach first runs a binary-attention based named entity recognition model to identify the raw paragraph's terms semantically and grammatically related to promotional content as suitable revision. 
The terms suitable for revision are then labeled with different revision entity categories (i.e., \textit{replacement} and \textit{insertion}) while the rest are categorized as \textit{unsuitability}. Finally, the promotional content is added to the raw paragraph by either inserting after or replacing the term, based on the term's revision entity\ignore{ category}.
%

\noindent\textbf{Revision entity categories}.
\label{subsubsec:entityCategory}
Here we present the revision entity categories\ignore{ (i.e., replacement, insertion)} and the way to ensure language smoothness and grammatical correctness using these categories. 


%

Specifically, for the replacement entity\ignore{ category}, \ignore{the }replaced terms\ignore{ or phrases} should be semantically and grammatically similar to \ignore{the }promotional content. For example, given the promotional content ``ABC\ignore{XL} Pharmacy''\ignore{ (a phishing-related online pharmacy)}, the drug description ``Rifaximin is ...\ignore{ an oral rifamycin} marketed \ignore{in the US }by Salix Pharmaceuticals'' contains semantically similar component ``Salix Pharmaceuticals'' that promotional content can replace, and also keeps grammatical correctness. 

The terms of the insertion entity play the role of antecedents to the promotional content to be injected and, therefore, should have proper semantic and grammatical relations with the promotional content to achieve contextual smoothness and grammatical correctness after insertion.
For example, the drug introduction ``Sofosbuvir, sold under the brand name Sovaldi among others, is a medication ...\ignore{ used to treat hepatitis C}'' contains a verb ``sold'' (a common drug promotional keyword~\cite{wangdemystifying}), which can be followed by an adverb phrase involving the promotional content: ``in ABC Pharmacy'', so as to ensure grammatical correctness and language smoothness.   
Also, we found that the target query term, once appearing in the raw paragraph, tends to be a suitable antecedent for the promotional content in the form of an adverb phrase.

 \noindent\textbf{Binary-attention based named entity recognition}.
Different from state-of-the-art tasks of named entity recognition (i.e., NER)~\cite{chen2022lightner,hu-etal-2022-adaptive}, our entity
recognition is expected to consider the relations between each term in the raw paragraph and the promotional content and target query. The aim is to uphold the quality of semantics and grammar when injecting promotional content. This is achieved in our study using a model built on top of the binary-attention based BiLSTM-CRF (the binary-attention based bidirectional LSTM and Conditional Random Fields). 
BiLSTM-CRF is the most commonly-used architecture for NER due to its lightweight design and high accuracy~\cite{li2022survey,yan2019tener}.
Improved from BiLSTM-CRF, this model uses deep attention mechanisms to capture the entity's attention on the raw paragraph context, promotional content, and target query.

The framework, as depicted in Fig.~\ref{fig:promotionBoosting}, takes the promotional content, target query, and raw paragraph as inputs, and produces an entity sequence for all terms in the raw paragraph. It labels each term of the raw paragraph sequentially.
Specifically, when the model labels a term $t$ in the raw paragraph, it first employs BiLSTM to encode the inputs. 
Next, to capture the relations of the raw paragraph with the promotional content and query (called outer information), we use an outer attention layer, which is based on the attention mechanism~\cite{bahdanau2015neural}. This layer computes the relation between $t$ and the terms of the outer information, resulting in the outer information vector.
Also, to strengthen the learning of semantic and grammatical context, we use a contextual attention layer based on the self-attention mechanism~\cite{vaswani2017attention}. This layer calculates the relation of $t$ to other terms in the raw paragraph and gets the context vector.
Finally, we concatenate $t$'s encoding vector with its outer information vector and context vector. The resulting vector is then fed into CRF, which outputs the probabilities of entities. We label $t$ with the entity that has the highest probability.
From all the raw paragraph terms suitable for revision, we randomly select a term to add promotional content to, based on its revision entity category. 

\subsection{Retrieval \& Insertion of Promotion Paragraph \ignore{Multi-task Adversarial Passage Retrieval\ignore{ Model}}}
\label{subsec:multitaskRetrieval}

\begin{figure}[t!]
\centering
\includegraphics[width=8.8cm]{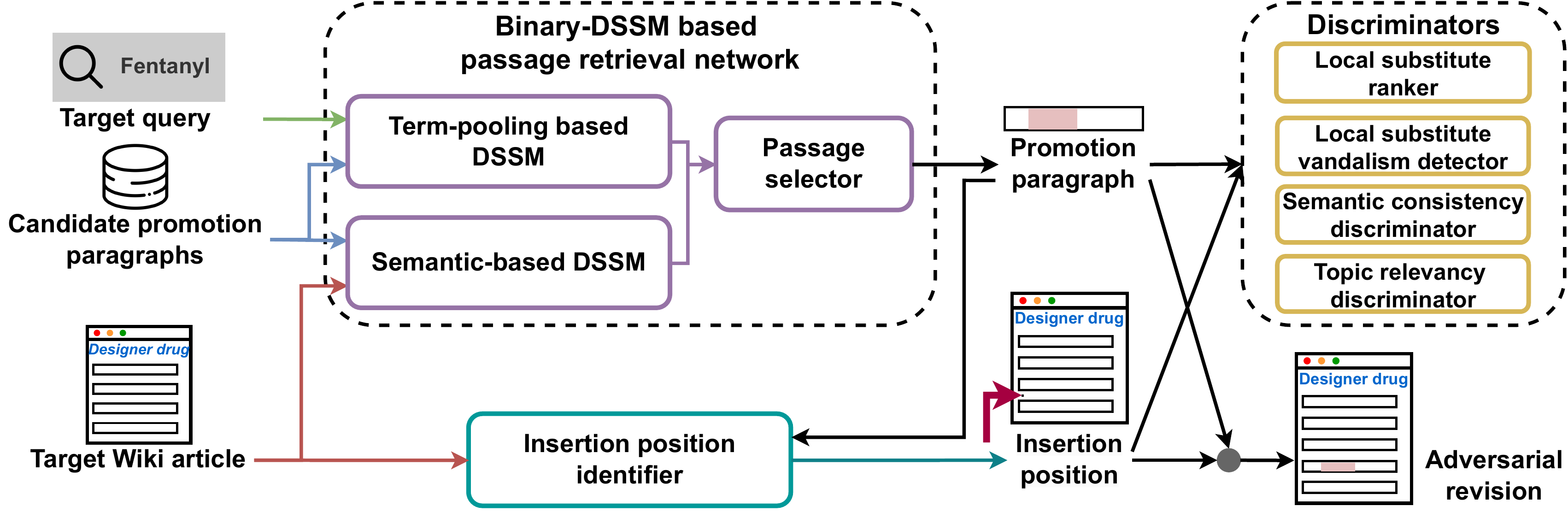}
\caption{Architecture of multi-task adversarial passage retrieval model.}
\label{fig:retrieval_model}
\end{figure}

\noindent\textbf{\boxed{Challenges}}
There are two main challenges when retrieving a promotion paragraph from the set of candidate promotion paragraphs and inserting it for adversarial revisions in a Wiki article. 
The first challenge is to simultaneously satisfy four attack objectives: rank boosting, detection evasion, semantic consistency, and topic relevancy. These objectives are crucial for successful adversarial revisions, as explained in Section~\ref{subsec:threatmodel}. 
Second, it is difficult to disorder search ranking and evade vandalism detection in the black-box setting, as adversaries are unaware of the model structures of Wiki search engine and vandalism detection tools.  


\noindent\textbf{\boxed{Our solution}}
Aiming for the successful illicit business promotion in Wiki systems, the adversary would insert into the target Wiki article a suitable promotion paragraph so that the revision satisfies the attack objectives. Thus, we propose a \textbf{\textit{multi-task adversarial passage retrieval model}} (see Fig.~\ref{fig:retrieval_model}) to find suitable promotion paragraphs, which can disorder article ranking while ensuring evasion of vandalism detection, semantic consistency, and topic relevancy.
Our approach includes a novel passage retrieval network (which retrieves a paragraph from a set of candidate promotion paragraphs tailored by the incentive injection model), an insertion position identifier, and four discriminators (i.e., a local substitute ranker, a local substitute vandalism detector, a topic relevancy discriminator, and a semantic consistency discriminator). These discriminators help optimize the passage retrieval network based on a trade-off among different training objectives.
To uncover the vulnerabilities in black-box Wiki search engine and vandalism detector, we used the knowledge distillation method~\cite{hinton2015distilling} to approximate these models and achieved substitute models as discriminators, such as the local substitute ranker and local substitute vandalism detector.
Here we discuss the details of these components and the training procedure.

\noindent\textbf{Promotional paragraph retrieval}. 
To retrieve the high-quality promotion paragraph in terms of relevancy, consistency, and ranking-prioritization, we build a passage retrieval network with Deep Structured Semantic Model (i.e., DSSM), a practical supervised text mining framework. DSSM maps the representations of both the query and the document onto a semantic space and computes their similarity~\cite{huang2013learning}. Since our retrieval model utilizes both the article's semantic information and the query's word-semantic mixed information, different inputs (i.e., target query and target article) end up having different semantic spaces. To address this issue, we propose a binary-DSSM based passage retrieval network, in which each input has a distinct DSSM delineation\ignore{the DSSM for each input has a different delineation} and the two DSSMs' outputs (the similarity between each input and a given candidate promotion paragraph) are then combined to provide an overall relevance for the candidate promotion paragraph.
Below we elaborate on the parts of our passage retrieval network.

\noindent$\bullet$\textit{ Binary-DSSM}. 
Since both dependencies between the paragraphs $\{p\}$ and the target query $q$, as well as the target article $a$, contain essential information for retrieval, we develop a binary-DSSM based passage retrieval network to use this information. This network combines a traditional semantic-based DSSM~\cite{huang2013learning} with a novel term-pooling based DSSM.
Two DSSMs ensure the relevance of the retrieved paragraph to both the target article and the target query, respectively.

The term-pooling based DSSM or simply \textit{TermPool-DSSM} is newly proposed to capture the relevance between a query and a candidate promotion paragraph, based upon the combination of their semantic similarity complemented by term matching statistics, as utilized by real-world state-of-the-art search engines to determine the relevance of search results (like Google~\cite{exactKW,semanticseo\ignore{,kwImportance}}, Bing~\cite{exactKWBing}, Yahoo~\cite{exactKWYahoo}, etc.).

Specifically, TermPool-DSSM first calculates the semantic similarity between the semantic representations of the target query ${v_s}^q$ and one candidate promotion paragraph ${v_s}^p$, in the same way as the semantic-based DSSM~\cite{huang2013learning}.

To capture the term matching statistics in the form of \textit{word-density similarity}\ignore{~\cite{shen2014learning,palangi2016deep}}, TermPool-DSSM further computes the word-density similarity score between the word sequence of the query $\{{w}^q\}$ and that of this paragraph $\{{w}^p\}$.
For this purpose, we first use a fully-connected network to encode the pre-trained embeddings of each query word and each paragraph word, achieving word vectors $\{{v_w}^q\}$ and $\{{v_w}^p\}$, respectively. Then, a similarity score is calculated between ${v_w}^q$ and ${v_w}^p$ by the cosine distance. 
These similarity scores, which constitute the term statistic information, are then used to compute the word-density similarity.

A challenge here is similarity score sparsity: most ($w^q$, $w^p$) pairs have low similarities, so they contribute little to determining the relevance between the query and paragraph. 
To address this issue, we apply max pooling~\cite{lecun1998gradient} and $k$-max pooling~\cite{kalchbrenner2014convolutional} to extract the maximum matching statistics (the highest similarity score) and the matching density statistics (the average similarity score for top-$k$ pairs) of each query word, respectively. 
Then, each query word's similarity score for the paragraph is calculated as the mean of its maximum score from max pooling and its matching density score from $k$-max pooling.
This way, we can focus on the most valuable information, ignoring other less meaningful statistics (the pairs with low similarities). 
%
%
Finally, the word-density similarity score between a query and a paragraph is calculated by summing each query word's similarity score, indicating the query's term matching density in the paragraph.

Based upon the semantic similarity and term matching density between the query and the candidate promotion paragraph, we can compute the similarity $\mathrm{Sim}^q$ of this paragraph to the query via simply multiplying the semantic similarity by the word-density similarity score. 
This similarity will further be combined with the similarity $\mathrm{Sim}^a$ between the target article and this paragraph, computed by the semantic-based DSSM model, to determine the overall relevance of this paragraph to the inputs of the retrieval model.

\noindent$\bullet$\textit{ Passage selector}. The final layer in our retrieval network computes the posterior probability of retrieving a candidate promotion paragraph from $\{p\}$, the set with $n$ paragraphs.
With the similarities $\{\mathrm{Sim}\}\in\mathbb{R}^{n}$ output by each sub-model (TermPool-DSSM and semantic-based DSSM) across all candidate promotion paragraphs, we get a pair of continuous logit vectors, one for the query and the other for the article. 
Since the retrieval is guided by two inputs $\{q, a\}$, the posterior probability distribution of candidate promotion paragraphs being relevant is calculated by first running a softmax function on each continuous logit vector and then adding the output vectors together as defined in Equation~\ref{equ:final_prob2}. 
\begin{equation}
\footnotesize
    P(\{p\}) = \text{\sc Softmax}(\{{\mathrm{Sim}}^{q}\}_{\{p\}}) + \text{\sc Softmax}(\{{\mathrm{Sim}}^{a}\}_{\{p\}})
\label{equ:final_prob2}
\end{equation}

During inference, our passage retrieval network returns the paragraph $\overline{p}$ with the highest probability in the distribution as the retrieved paragraph. 
During training, to locate the paragraph not only relevant (to $q$ and $a$) but also best suited to meet four attack objectives, our network returns the retrieved paragraph representation $\widetilde{p}$, summing the top-$k$ paragraphs' word/semantic vectors weighted by their normalized probabilities, as the input of four discriminators.

\noindent\textbf{Insertion position identification}. 
To let the inserted paragraph be semantically smooth with the neighboring paragraphs $\{p^a\}$ of the target article, it is crucial to identify a suitable insertion position in the context of the article.
%
To this end, we calculate the mean cosine similarity between the retrieved paragraph\ignore{ (i.e., $\widetilde{p}$ during training and $\overline{p}$ during inference)} and each pair of neighboring paragraphs in the article. We then utilize a softmax function to convert the sequence of similarities into an output vector. From this vector, we select the pair of paragraphs $p^a_I$ and $p^a_{I+1}$ with the highest probability, which is the optimal insertion position.

\noindent\textbf{Discrimination for retrieval and insertion}. 
\label{subsubsec:discriminator}
We adopt the generative adversarial networks (GANs) design~\cite{goodfellow2020generative} to train the passage retrieval network in conjunction with multiple discriminators associated with four attack objectives (see Section~\ref{subsubsec:architecture}). Below we elaborated on each discriminator.

\noindent$\bullet$\textit{ Local substitute ranker}.
\label{subsubsec:rankApprox}
To identify the paragraph helping boost a Wiki article's rank, we utilize adversarial learning by training the retrieval network against the target search engine\ignore{ranking system}.
Note that our method treats Wiki search engine as a black-box system to make our attack more general.

Specifically, we construct a local substitute model approximating the target black-box search engine\ignore{ranking system}. This substitute model is based on the learning-to-rank model~\cite{liu2009learning} and serves as a discriminator in adversarial learning. 
Prior research~\cite{he2018adversarial} shows that a small learning-to-rank model with limited ability is more robust. 
So, in our study, we chose MV-LSTM~\cite{wan2016deep} (see Appendix~\ref{app:submodels}) as the small pointwise learning-to-rank substitute model.
%
%
When training the passage retrieval model, the substitute ranker takes word vector sequences of a target query $q$ and a revised target article $\widetilde{a}$ as inputs.
Its output is a pseudo-ranking score as follows: 
\begin{equation}
\footnotesize
    \mathrm{Score}(\widetilde{a}|q) = \text{\sc Ranker} (q, a \oplus \widetilde{p})
\label{equ:subrankscore}
\end{equation}
where $\oplus$ denotes injection, i.e., concatenating the retrieved paragraph representation $\widetilde{p}$ with the target article context $a$. To find the candidate promotion paragraph that can improve the target article's rank, 
the adversarial loss is set to:
\begin{equation}
\footnotesize
    \mathcal{L}_{\mathrm{rank}} = - \mathrm{Score}(\widetilde{a}|q)
\label{equ:subrankloss}
\end{equation}

\noindent$\bullet$\textit{ Local substitute vandalism detector}.
\label{subsubsec:vandalApprox}
For a stealthy attack, our edits aim to evade Wiki's automatic vandalism detection.
Since vandalism detectors are considered black-box too, we train a local substitute to approximate the target model.

To this end, we build a local substitute vandalism detector that analyzes both the edit itself and the edit under the context of the target article.
Specifically, we use a Transformer~\cite{vaswani2017attention} to encode the edit information in the retrieved paragraph representation $\widetilde{p}$, and an Enhanced Sequential Inference Model~\cite{chen2016enhanced} to collect the text-matching information between $\widetilde{p}$ and the target article ${a}$. 
These models produce two vectors, which are concatenated and fed into a fully-connected network to calculate a binary pseudo-damaging probability $(d_{\mathrm{True}}, d_{\mathrm{False}})$.
To evade detection of the substitute detector, the pseudo-damaging probability should have a low probability of ``True'' and a high probability of ``False'', trained by the cross-entropy loss:
\begin{equation}
\footnotesize
    \mathcal{L}_\mathrm{detect} = \log(d_\mathrm{True}) - \log(d_\mathrm{False})
\label{equ:subvandloss}
\end{equation}

\noindent$\bullet$\textit{ Topic relevancy discriminator and semantic consistency discriminator}.
To ensure that the retrieved promotion paragraph is relevant to the target article's topic and maintains semantic consistency with its context, we include a topic relevancy discriminator and a semantic consistency discriminator in our approach. These discriminators evaluate the semantic similarities of the retrieved paragraph to the target article's topic and its neighboring context, respectively.

The topic relevancy discriminator uses cosine similarity loss to quantify the distance between the retrieved paragraph and the article topic (i.e., the Wiki article’s lead paragraph): 
\begin{equation}
\footnotesize
    {\mathcal{L}_\mathrm{topic}}(\widetilde{p}, p_\mathrm{topic}^a) = - \text{\sc Cosine}(\widetilde{p}, p_\mathrm{topic}^a)
\label{equ:globalloss}
\end{equation}

The semantic consistency discriminator extracts two neighboring paragraphs (i.e., $p_{I}^a$ and $p_{I+1}^a$) around the insertion position in the target article and computes the loss: 
\begin{equation}
\footnotesize
    {\mathcal{L}_\mathrm{sem}}(\widetilde{p}, \{p^a\}) = - 0.5 \cdot [\text{\sc Cosine}(\widetilde{p}, p_{I}^a) + \text{\sc Cosine}(\widetilde{p}, p_{I+1}^a)]
\label{equ:localloss}
\end{equation}

\noindent\textbf{Model training}. 
The multi-task adversarial retrieval model is trained by minimizing the weighted sum of four losses:
%
\begin{equation}
\footnotesize
    {
    \mathcal{L} = w_\mathrm{rank} \mathcal{L}_\mathrm{rank} + w_\mathrm{detect} \mathcal{L}_\mathrm{detect} + w_\mathrm{topic} \mathcal{L}_\mathrm{topic} + w_\mathrm{sem} \mathcal{L}_\mathrm{sem}
    }
\label{equ:loss}
\end{equation}
As fixed weights may not achieve the optimal balance among various objectives, the weights are optimized in each iteration by the Multiple Gradient Descent Algorithm~\cite{sener2018multi}. This algorithm uses a Franke-Wolfe optimizer to calculate the weight of each task based on its loss and gradient~\cite{sener2018multi}.

\section{Impelmentation and Evaluation}
\label{sec:evaluation}

\subsection{Implementation}
\label{subsec:implementation}
Here we briefly describe the implementation of a local victim Wiki system and our prototype system of MAWSEO.

\noindent\textbf{Local Wiki system}.
We deployed our local Wiki system utilizing \texttt{MediaWiki}~\cite{mediawiki} and installed the extensions relevant to our task, including \texttt{ORES}~\cite{oresIntr} for detecting vandalism and \texttt{CirrusSearch}~\cite{CirrusSearchScoring,helpCirrusSearch} for enhancing search. These two open-source extensions are widely installed in MediaWiki-based Wiki systems like Wikipedia and Wikidata~\cite{oresIntr,helpCirrusSearch}. The system was hosted on an Ubuntu PC with an Intel i7\ignore{ 1.99GHz} CPU and 16GB memory. 
For illicit online pharmacy promotion, we crawled 27,410 articles from Wikipedia under the Drug category with a depth of five, using them as the content of the local Wiki system. To enrich the corpus of other categories, we randomly collected 95,761 articles from Wikipedia for our system. In total, our local Wiki system consists of 123,171 articles and their respective edit histories on Wikipedia.

In our study, we used an editor account to conduct adversarial revision attacks on our Wiki system.
We used a typical account setting for a misinformation distributor~\cite{kumar2016disinformation} to mimic a real-world attack. Also, we set the account in the user group with low account privilege to trigger the detection of vandalism detectors for each revision. 
Specifically, we set editor accounts to be 45 days old~\cite{kumar2016disinformation} and in the user group with the lowest privilege, i.e., \textit{Registered (new) users}. In the user group of \textit{Registered (new) users}, each revision will trigger the test of vandalism detectors~\cite{UserAccessLevel}. Note that, in this user group, an editor account usually has no edit history.

\noindent\textbf{MAWSEO}.
In our implementation, we constructed our prototype system using \texttt{TensorFlow}~\cite{TensorFlow}. We ran our MAWSEO model on a Linux server with an Intel 6248 \ignore{2.5GHz 20-core }CPU, an NVIDIA V100-PCIE-32GB GPU, and 512 GB memory.

\noindent$\bullet$\textit{ Incentive injection model}. 
\label{subsubsec:injectiondata}
To build the dataset for incentive injection\ignore{ for an incentive injection model}, we annotated 1,030 paragraphs as the ground truth (80\% as the training set and the remaining as the test set). 
We evaluated the model on the test set, achieving a precision of 93.67\%, a recall of 95.00\%, and an F1 score of 94.33\%.

\noindent$\bullet$\textit{ Binary-DSSM based passage retrieval network}. 
Our network ran 40 epochs of training, with a batch size of 256 and a learning rate of 0.002. The pre-trained word embeddings were generated by GloVe~\cite{pennington2014glove}\ignore{, a pre-trained word vector set}. For generating the semantic representations of texts, we used SBERT~\cite{reimers-2019-sentence-bert}\ignore{, a pre-trained model for paragraph embeddings}.

\noindent$\bullet$\textit{ Local substitute ranker}.
To train the local substitute ranker\ignore{ based on MV-LSTM (see Section~\ref{subsubsec:rankApprox})}, we built the ranking dataset comprising both a training set and a test set, using the search results and their respective ranking scores obtained by querying our local Wiki system.
This model was evaluated on the ranking test set, achieving a Mean Square Error~\cite{taylor2008softrank} of 4.59 in ranking scores, as well as Normalized Discounted Cumulative Gains~\cite{jarvelin2017ir} of 96.43\% and 98.68\% for top-20 and top-200 search results. 

\noindent$\bullet$\textit{ Local substitute vandalism detector}.\label{subsubsec:vandalismData}
We constructed a vandalism dataset with 39,062 ORES \texttt{damaging}-labeled revisions to train the local substitute vandalism detector. The model was trained on an 80\% random sample from this dataset and evaluated on the remaining 20\%, achieving 87.29\% precision, 85.33\% recall, and an 86.30\% F1 score.


More details on the sub-model's dataset and evaluation can be found in Appendixes~\ref{app:submodels}.

\subsection{Experiment Setup}
\noindent\textbf{Dataset}.
\label{subsubsec:dataset}
To evaluate MAWSEO, we ran our prototype on the following datasets.
 
\noindent$\bullet$\textit{ Promotional content and target queries}.
For illicit online pharmacy promotion, we used 125 illicit online pharmacies in the ConcoctedPharma dataset~\cite{ConcoctedPharma} as the target businesses to be promoted. 
We randomly assigned one target business to pollute the search results of several specific queries.

We generated 659 query terms based on the most popular terms in Wikipedia associated with drug names. More specifically, we used popular medical topics in Wikipedia~\cite{wikiPopMedicalPages} and chose those related to drug names as the queries. The average length of the query terms is 1.1.
After that, we collected all search results of those queries. 
In the evaluation, we divided the search results into two sets: top-20 (i.e., the first page) search result set $D^{(\mathrm{t20})}$ and sampled all search result set $D^{(\mathrm{all})}$\ignore{ (i.e., top-1000 search result set)}.
Particularly, for $D^{(\mathrm{all})}$, we randomly sampled ten search results from the search results of each rank range of 100 (i.e., top 100, top 100-200, etc.).
In this way, we obtained 31,460 query-article pairs for $D^{(\mathrm{all})}$, which are 12,778 query-article pairs for $D^{(\mathrm{t20})}$.
Note that, to train the binary-DSSM based passage retrieval network, we randomly selected 527 (80\%) query terms and their associated 25,675 query-article pairs for training ($D_\mathrm{train}^{(\mathrm{all})}$) and the rest for testing ($D_\mathrm{test}^{(\mathrm{all})}$).

\noindent$\bullet$\textit{ WikiDump dataset}.
As previously noted in Section~\ref{subsec:overview}, the raw paragraphs utilized in our study were sourced from WikiDump, containing a vast collection of over 21 million Wiki articles. To run the prototype efficiently, we selected 53,826 paragraphs pertaining to medical subjects as raw paragraphs.
When contaminated with different promotional content, 76.86\% of the raw paragraphs can be successfully polluted by the incentive injection model on average.


\noindent\textbf{Baseline Approaches}.
\label{subsubsec:baselines}
We implemented three language-generation based adversarial ranking attacks \textit{HotFlip}~\cite{ebrahimi2018hotflip,song2020adversarial}, \textit{Collision}~\cite{song2020adversarial}, and \textit{PAT}~\cite{liu2022order} for comparison. We elaborated the design of these three models in Appendix~\ref{app:baselineDesign}.


\begin{table*}[!t]
\scriptsize
\centering
\caption{MAWSEO performance on different search result sets}
\label{table:promotionResult}  
\begin{tabular}{l|l|c|c|c|c|c|c}
\hline

\makecell{Search Result Set} & Approach & \makecell{\% Rank\\Boosting Succ.} & \makecell{\% Evasion\\Succ.} & \makecell{\% Topic\\Relevancy} & \makecell{\% Semantic\\Consistency} & \makecell{\% Promotion\\Succ.} & \makecell{Time Cost\\(hrs)}\ignore{\makecell{Ave.\\Time (s)}} \\ 
\hline
\hline
\multirow{4}*{\makecell{Top-20 ($D^{(\mathrm{t20})}$)}}  & \textbf{MAWSEO} & \textbf{46.42} & \textbf{81.81} & \textbf{81.47} & \textbf{78.99} & \textbf{28.09} & \textbf{38.64\ignore{0.14}} \\
\cline{2-8}
 & HotFlip & 28.66 & 57.69 & 4.65 & 0.24 & 0.02 & 178.89\ignore{50.40} \\
\cline{2-8}
 & Collision & 26.82 & 41.87 & 5.25 & 6.15 & 0.20 & 532.84\ignore{150.12} \\
 \cline{2-8}
 & PAT & 25.67 & 65.35 & 0.52 & 0.05 & 0.01 & 950.90\ignore{267.90}\\
\hline
\multirow{4}*{All ($D_\mathrm{test}^{(\mathrm{all})}$)} & \textbf{MAWSEO} & \textbf{68.37} & \textbf{91.50} & \textbf{59.72} & \textbf{52.45} & \textbf{30.27} & \textbf{36.99\ignore{0.24}} \\
\cline{2-8}
 & HotFlip & 27.78 & 76.21 & 3.18 & 0.40 & 0.02 & 76.89\ignore{47.85}  \\
\cline{2-8}
 & Collision & 27.35 & 60.47 & 42.78 & 10.16 & 1.11 & 197.22\ignore{122.73} \\
 \cline{2-8}
& PAT & 27.61 & 81.21 & 1.07 & 0.16 & 0.03 & 397.70\ignore{247.49} \\
\hline
\end{tabular}
\end{table*}

\noindent\textbf{Metrics}.
We measured MAWSEO using the below metrics.

\noindent$\bullet$\textit{ Rank boosting success rate}. To measure ranking manipulation, we defined the rank boosting success rate as the number of revisions that boost articles' ranks divided by the total revision number.
Our evaluation first focused on the first page of search results, an important target for cybercriminals. To assess our model's generalization in ranking manipulation, we also measured its performance on all search results.

\noindent$\bullet$\textit{ Evasion success rate}. The evasion success rate measures the effectiveness of MAWSEO revisions in evading automatic vandalism detection. It is calculated by dividing the number of revisions undetected as vandalism\ignore{ that are not detected as vandalism by the target detector} by the total number of revision samples.

\noindent$\bullet$\textit{ Topic relevancy rate and semantic consistency rate}. We used cosine similarity to measure semantic similarity between paragraphs encoded by SBERT~\cite{reimers-2019-sentence-bert}. 
The topic relevancy and semantic consistency rates measure how many revisions preserve semantic relationships between the promotion paragraph with the article topic (i.e., the Wiki article's lead paragraph) and two neighboring paragraphs, respectively. 
The criterion of preserving topic relevancy is whether the semantic similarity between the promotion paragraph and article topic (i.e., topic similarity) exceeds a topic threshold, and that of maintaining semantic consistency is whether the average similarity between the promotion paragraph and its two neighboring paragraphs (i.e., neighboring similarity) exceeds a consistency threshold. 
%
We computed average topic and neighboring similarities from 15,000 Wiki articles, yielding thresholds of 0.33 and 0.39, respectively.

\noindent$\bullet$\textit{ Promotion success rate}. We defined a revision as an adversarial revision for illicit promotion on the Wiki systems when a revision satisfies all the attack objectives in Section~\ref{subsec:threatmodel}.
%
The promotion success rate is the percentage of adversarial revision samples that successfully meet all attack objectives out of all the revisions.

\subsection{End-to-end Effectiveness and Efficiency\ignore{Efficacy}} 
\label{subsec:promotionEffective}
We evaluated the effectiveness of MAWSEO on the top-20 and all search results (i.e., $D^{(\mathrm{t20})}$ and $D^{(\mathrm{all})}_\mathrm{test}$) of target queries.
%
%
Table~\ref{table:promotionResult} summarizes the evaluation results and compares them with the baseline approaches. 
MAWSEO produced 3,589 (28.09\%) and 1,751 (30.27\%) adversarial revision samples satisfying all attack objectives on the top-20 and all search results, taking 38.64 and 36.99 hours, respectively. 
Also, compared to baseline approaches, our retrieval-based method was much more effective and efficient.
%
HotFlip generated just 2 and 1, Collision produced 26 and 64, and PAT yielded only 1 and 2 adversarial revisions that met all attack objectives for the top-20 and all search results, respectively.
Sampled revisions by MAWSEO and baseline approaches are presented on the \href{https://sites.google.com/view/mawseo}{project website}.

\begin{table}[!t]
\scriptsize
\centering
\caption{Ranking manipulation in different rank levels}
\label{table:rankonlevel}  
\begin{tabular}{l|c|c}
\hline
Rank Level & \makecell{Ave. Ranking\\Manipulation Margin ($\uparrow$)} & \makecell{\% Rank Boosting Succ.} \\ 
\hline
\hline
2-100 & 7.37 & 53.29\\
\hline
101-200 & 30.66 & 67.53\\
\hline
201-300 & 61.38 & 75.63 \\
\hline
301-400 & 76.61 & 77.07\\
\hline
401-500 & 122.25 & 82.37 \\
\hline
\end{tabular}
\end{table}

\noindent\textbf{Ranking manipulation}.
We further evaluated ranking manipulation at various rank levels (i.e., top 100, top 100-200, etc.).
Table~\ref{table:rankonlevel} shows that, in total, the average ranking manipulation margin of MAWSEO revisions was 152.51, which was 7.37 for top 2-100, 30.66 for top 101-200, and 61.38 for top 201-300. 
Among revisions targeting ranks 21-100, 24.49\% revisions per query appeared on average in the top 20 (first page of search engine results). 
31.50\% revisions with the rank level of top 101-200 appeared in the top 100 results.
Notably, MAWSEO has a greater impact on lower-ranking articles, resulting in a higher success rate and larger manipulation margin. For instance, articles\ignore{ that were originally} ranked between 900 and 1000 saw an average increase of 279.17 places with an 87.69\% success rate, surpassing\ignore{ the success rates and margins achieved by articles initially} those ranked higher.

\noindent\textbf{Additional vandalism detection\ignore{detectors}}.
In the study, we calculated the evasion success rate based on the state-of-the-art vandalism detector, ORES \texttt{damaging} detection model (see Sections~\ref{subsec:threatmodel}). Here, we also evaluated the vandalism detection evasiveness of MAWSEO-generated revisions against four other vandalism detection models. These four detectors include two less-used vandalism detectors of ORES (i.e., ORES \texttt{goodfaith} and ORES \texttt{revert}~\cite{oresMediaWiki}), both of which detect vandalism behaviors based on different intents and different models as well as are trained by different datasets, and two third-party vandalism detectors (i.e., ClueBot NG~\cite{cluebotng} and AVBOT~\cite{AVBOTblog,AVBOTwiki}). These four detectors are not considered in the training of MAWSEO.
%
When detecting the revisions crafted from $D^{(\mathrm{t20})}$ and $D_\mathrm{test}^{(\mathrm{all})}$, evasion success rates listed in Table~\ref{table:additionalvand} indicate that the MAWSEO-generated revisions are highly evasive against real-world vandalism detectors, which all consider the content and language of edits in detection.
Looking into some revisions detected by those vandalism detectors,
some revisions detected due to blocklist words could yield high false positives. 
%
For example, the revision on the article ``\textit{Botryosphaeria disrupta\ignore{Phytophthora cinnamomi}}'' was detected as vandalism by AVBOT due to the appearance of the blocklist words ``amazing'' and ``shit'' (set by AVBOT) in the inserted paragraph. Note that this revision can bypass the ORES detection models and ClueBot NG.

\begin{table}[!t]
\scriptsize
\centering
\caption{Evasion success rates against \ignore{additional }vandalism detectors}
\label{table:additionalvand}  
\begin{tabular}{l|c|c}
\hline
Model & Top-20 ($D^{(\mathrm{t20})}$) & All ($D^{(\mathrm{all})}_\mathrm{test}$)  \\
\hline
\hline
ORES \texttt{damaging} & 81.81\% & 91.50\%  \\
\hline
ORES \texttt{goodfaith} & 99.94\% & 99.91\% \\
\hline
ORES \texttt{revert} & 100\% &  100\% \\
\hline
ClueBot NG & 99.78\% & 99.79\%  \\
\hline
AVBOT & 100\% & 99.97\%  \\
\hline
\end{tabular}
\end{table}

\subsection{User Reachability and Evasiveness}
\label{subsec:userstudy}
In this study, we aim at answering the following three research questions: \textbf{Q1}: whether the Wiki system's normal users can correctly identify the Wiki articles that were modified by MAWSEO? \textbf{Q2}: whether the promotional content was delivered to normal users?
\textbf{Q3}: Whether revisions can bypass the vandalism reporting of potential content moderators?
The study is conducted with our institution’s IRB approval (see Section~\ref{subsec:ethical}).

\noindent\textbf{Recruitment}.
This user study\footnote{More user study details, including the questionnaire sample, validation criteria, and further evaluation, are \ignore{elaborated }in Appendix~\ref{subsec:invalidResponse} and the \href{https://sites.google.com/view/mawseo}{project website}.} was conducted through Amazon Mechanical Turk (MTurk). We recruited adult participants living in the U.S. who could read and write in English. Each participant will receive \$2.
After removing 18 invalid responses, we totally collected 104 valid responses with diverse backgrounds: ages range from 18 to 54+ (39\% are female and 61\% are male); education ranges from high school to graduate degree; 15 various categories of occupation. 97\% participants have known Wikipedia for over three years, and all participants have read articles from it. 65\% read Wikipedia articles a few times per week or more frequently, and 25\% read a few times per month.

\noindent\textbf{User awareness of malicious modification}. 
The survey is designed to ask participants to what extent they think the modified information appears in Wiki articles. Specifically, we randomly selected 30 articles (referred to as malicious articles) generated by MAWSEO, along with 30 randomly selected Wiki articles (benign articles) from Wikipedia. 
We also compared MAWSEO with three baseline approaches: HotFlip, Collision, and PAT. To this end, we generated 45 articles using HotFlip, Collision, and PAT (15 for each baseline). Totally, our pool contains 105 unique articles.

In the survey, participants were asked to read four articles, randomly chosen from the article pool, presented in a UI designed to mimic the Wikipedia page format for an authentic reading experience and atmosphere.
%
%
For each article, we asked subjects whether they thought the article was trustworthy. Next, if subjects selected ``no'' or ``sort of'',
we asked them why untrustworthiness (an open-ended question) and whether they found the misinformation problem (i.e., containing fabricated content, false context of connection, etc.). After that, we asked subjects to choose all problematic paragraphs that make the article untrustworthy.

In total, we received 128 valid answers to malicious articles, 176 valid answers to baseline articles, and 112 valid answers to benign articles from 104 subjects.
From the survey, 120 (94\%) malicious articles were considered to be trustworthy, compared to 103 (92\%) in benign articles and 44 (25\%) in baseline articles. 
It shows that the subjects read articles carefully and had the ability to tell the quality of the articles. More importantly, our survey result indicates that the articles modified by MAWSEO have high trustworthiness like those benign articles. Thus, our malicious articles have the potential ability to bypass the Wiki users' awareness of malicious modification.
Most of the subjects who expressed distrust toward the remaining 6\% of malicious articles mentioned that they were unfamiliar with the medical topics and slang terms: e.g., ``I can't say for sure that it's trustworthy because I don't know enough about the topic. I don't believe everything that I read on the internet when it comes to drugs/pharm things.''
Meanwhile, most (50\ignore{52}\%) untrustworthy articles were ascribed by subjects to neither fabricated content nor false context of connection (misinformation types), indicating that misinformation we inserted cannot be identified by subjects in those malicious articles even if the articles have been identified as untrustworthiness. When we further required them to choose the paragraphs having problems like misinformation, only 3 (2\% in all malicious articles) modified paragraphs were correctly selected compared to 118 (67\%) in those baseline articles.

\noindent\textbf{User reachability of promotional content}. 
To understand the reachability of promotional content to Wiki users, we asked participants to select the topics covered by the articles. Specifically, we carefully devised a multi-choice question with four choices in the survey. Each choice represents a topic covered by a section in the article, with one of the choices implicitly covering the recapitulative topic of promotional content (e.g., ``CP-122 sold in an online drug store was found to be effective as a treatment for migraine").
To avoid the leakage of implicit hints by the question to participants, we presented the question after participants finished reading the article. If the promotional content topic appears in participants’ selected choices, we believe participants reached promotional content.
The result shows that 111 (87\%) promotional content texts in malicious articles were reached out by participants, compared to 71 (40\%) in baseline articles.
We concluded \ignore{from the participants' feedback }that Wiki users received promotional content and accepted it as part of the topics conveyed in malicious Wiki articles.
We further discussed whether the insertion position would affect user reachability and found no significant correlation between reachability and insertion positions (see the \href{https://sites.google.com/view/mawseo}{project website}).

\begin{table*}[!t]
\scriptsize
\centering
\caption{Performance of rank-boosting alternated MAWSEOs on different search result sets}
\label{table:alternative}  
\begin{tabular}{l|l|c|c|c|c|c|c}
\hline
Search Result Set & \makecell{Rank-boosting Alternated\\MAWSEO} & \makecell{\% Rank\\Boosting Succ.} & \makecell{\% Evasion\\Succ.} & \makecell{\% Topic\\Relevancy} & \makecell{\% Semantic\\Consistency} & \makecell{\% Promotion\\Succ.} & \makecell{Time Cost\\(hrs)}\ignore{\makecell{Ave.\\Time (s)}} \\ 
\hline
\hline
\multirow{2}*{Top-20 ($D^{(\mathrm{t20})}$)} & Keyword-stuffing & \ignore{46.42}53.52 & 79.97 & 72.93 & 69.53 & 26.19 & 38.64\ignore{0.15} \\
\cline{2-8}
~ & Synonym-substituting & 41.88 & 79.86 & 77.92 & 74.24 & 22.39 & 70.19\ignore{9.03} \\
\hline
\multirow{2}*{All ($D_\mathrm{test}^{(\mathrm{all})}$)} & Keyword-stuffing & \ignore{68.37}72.52 & 87.14 & 51.79 & 45.08 & 23.30 & 36.99\ignore{0.25} \\
\cline{2-8}
~ & Synonym-substituting & 58.17 & 88.94 & 57.34 & 47.29 & 22.26 & 59.41\ignore{14.19} \\
\hline
\end{tabular}
\end{table*}

\noindent\textbf{Reviewer acceptance of revision}. 
To study whether the generated revisions can bypass content moderation, we simulate the reviewer mode of Wiki systems and identify 30 participants with Wikipedia article review and revising experience (37 review revisions on average) for revision content moderation. 
More specifically, all participants were asked to read Wikipedia policies on review~\cite{wikireviewerguide} and vandalism~\cite{wikivandPolicy} before our study. When reviewing a pending revision, a participant will review a revision in the reviewer mode of Wiki systems, where a page shows the difference between the latest accepted revision and the new revision to the article~\cite{wikireviewer}.
%
%
%
The result shows that 29 (62\%) revisions on malicious articles were accepted, compared to 2 (4\%) on baseline articles. It indicates that our malicious articles have the potential ability to pass the vandalism checking of Wiki reviewers. 
Also, when we asked participants the reason why they selected ``no'' or ``need further modification'', 5 answers (from 56\% of the participants who do not accept changes) did not refer to misinformation (i.e., fabricated content and false context of connection). 
%

\subsection{Ablation Study}
\label{sec:abla}
The high and balanced promotion effectiveness of MAWSEO is the result of combining multiple tasks. To gain a better insight into how the four discriminators contribute to MAWSEO's effectiveness, we conducted an ablation study. This study focused on two aspects: (1) evaluating each discriminator's contribution to promotion, and (2) exploring the impact of functional module alternation by using keyword stuffing and adversarial synonym substitution as alternatives to the substitute ranker discriminator for rank boosting.


\noindent\textbf{Contribution of discriminators to end-to-end performance}.
\label{subsubsec:abla4integral}
We first evaluated the contribution of individual discriminators by comparing the performance of the models trained without each discriminator.
As shown in Table~\ref{table:remove1discriminator} of Appendix~\ref{subsec:ablation1Discriminator}, we observed that removing one target discriminator would lead to a drop in both results of the illicit promotion and this discriminator's associated sub-task, but an elevation in other sub-tasks' performance.
%

\ignore{We first evaluated the contribution of discriminators by comparing the performance of the models trained by such a discriminator's different loss weight values (see Equation~\ref{equ:loss}).
Compared with Table~\ref{table:promotionResult}, we observed that the decrease of one target discriminator's loss weight value would lead to a drop in the results of the illicit promotion and this discriminator's associated sub-task, while would render an elevation in other sub-tasks' performance.
On the contrary, a much higher loss weight value of one target discriminator would suppress the performance of other sub-tasks, causing the ineffectiveness in illicit promotion.
Also, \ignore{In addition, we found that }the change in the loss weight value of the topic relevancy discriminator or semantic consistency discriminator would lead to a change in both topic relevancy and semantic consistency due to the high dependency between these two discriminators.
According to the above study, we chose the\ignore{ loss weight} values with $w_\mathrm{rank}=0.1$, $w_\mathrm{detect}=0.3$, $w_\mathrm{topic}=2.5$, and $w_\mathrm{sem}=1$ for a good trade-off among all sub-tasks.
We summarize the experiment results on $D_\mathrm{test}^{(\mathrm{all})}$ and $D^{(\mathrm{t20})}$ in Fig.~\ref{fig:top1000+20Abla} and Fig.~\ref{fig:top20Abla} of Appendix~\ref{subsec:ablationgraph}. 
}

To assess each sub-task's upper-bound performance\ignore{ of each sub-task constrained by the associated discriminator}, we conducted uni-constrained tests, where we only retained one corresponding discriminator of the target sub-task. 
The results, shown in Appendix Table~\ref{table:uniconstrainedAttack_main}, reveal that the sub-task constrained by the retained discriminator outperforms the same sub-task with complete constraints. However, the uni-constraint model causes lower promotion effectiveness.

\noindent\textbf{Rank boosting alternative}.
To evaluate the efficacy of rank-boosting alternatives---like the blackhat SEO technique~\cite{liao2016characterizing} and the synonym substitution based adversarial ranking attack~\cite{wu2022prada} introduced in Section~\ref{subsec:promotionInRanksys}---in terms of disordering Wiki search for promotion, we conducted an ablation study.

\ignore{Given that traditional blackhat SEO~\cite{liao2016characterizing} practice and adversarial ranking attack~\cite{wu2022prada} have been used or newly proposed for rank manipulation against ranking systems, here we conducted an ablation study to understand the efficacy of these rank-boosting alternatives on Wiki illicit promotion.}

\noindent$\bullet$\textit{ Keyword stuffing}.
Keyword stuffing, a common blackhat SEO technique, is the repetition of keywords in an article to give more relevance to target query terms~\cite{liao2016characterizing}, to boost the article' rank. 
To compare MAWSEO with blackhat SEO, we chose this technique as an alternative for rank boosting. It is because, unlike other blackhat SEO techniques such as link farm spam~\cite{wu2005identifying} or sneaky redirection~\cite{leontiadis2011measuring}, keyword stuffing is based on changing the content of the article itself, not link building which is out of our study's scope. 

 
In our keyword-stuffing framework, MAWSEO is modified by removing the substitute ranker discriminator (see Section~\ref{subsubsec:rankApprox}) from the multi-task adversarial passage retrieval model.
%
Instead, we used the target query phrase as the keyword, and then randomly and repeatedly added it to the promotion paragraph\ignore{ until the article achieves a given keyword density}. 
This model is called keyword-stuffing MAWSEO. 
To ensure a fair comparison with MAWSEO, we adjusted the target article's keyword density $d=\frac{l \times f}{T}$ ($l$ is the keyword token number, $f$ is its frequency, and $T$ is the article token number)~\cite{liao2016characterizing} to achieve the same rank as that of MAWSEO. 
%
%
For example, if MAWSEO boosted the rank of the article ``\textit{Hemp}'' from 9 to 6 for a query ``\textit{CBD}'', we used the keyword stuffing technique to elevate this article to the same rank position (Rank 6) \ignore{, aligned with MAWSEO's rank boosting, }and led the article's keyword (query) density increases from 0.05\% to 0.27\%.
%
%

\noindent$\bullet$\textit{ Adversarial synonym substitution}.
PRADA is a newly-proposed method that employs synonym word substitution for adversarial ranking attacks~\cite{wu2022prada}. Since PRADA, similar to keyword stuffing, can not directly generate the promotion paragraph, we implement PRADA as another alternative for rank boosting.
In this framework, we still utilized the modified MAWSEO in keyword stuffing to get the promotion paragraph. 
Then, we applied PRADA to replace the important words in the paragraph with their synonyms. We name it synonym-substituting MAWSEO.


\noindent$\bullet$\textit{ Evaluation}.
%
%
Table~\ref{table:alternative} demonstrates MAWSEO's superiority over both keyword-stuffing and synonym-substituting MAWSEO. While keyword-stuffing MAWSEO increased average keyword densities in the top-20 and all search result articles to 1.32\% and 0.95\% from 0.56\% and 0.35\% respectively, yielding similar rank-boosting to MAWSEO, it produced fewer successful adversarial revisions (3,347 and 1,348) that meet the attack objectives. Likewise, synonym-substituting MAWSEO produced fewer successful revisions (2,861 and 1,288), indicating that neither keyword stuffing nor synonym substitution effectively replace the substitute ranker discriminator in MAWSEO for rank boosting.

\subsection{Revenue Analysis}
\label{subsec:revenue}
To understand potential practical impacts, we estimated the promotional revision's revenue. We used a revenue model in prior work~\cite{wangdemystifying}: $R(t) =V(t) \times r_{v} \times R_a$, where the total revenue $R(t)$ during a time period $t$ is calculated from the total number of post-impression actions taken (i.e., view-through number: the view-through rate $r_{v}$ times the number of views $V(t)$) and the average revenue per action $R_a$.



To estimate the view volumes $V(t)$ led by our ranking attack, we calculated the view counts of adversarial revisions based on their boosted ranks.
We calculated the average view volumes for articles with different ranks (1-1,000) based on the past-30-day views shown on Wikipedia~\cite{PageviewWiki}. Specifically, we collected the past-30-day views of articles with the same rank and calculated their average. In this way, we had the average view volume of an article in the past 30 days, with the rank of 1 being 49,467.8\ignore{31,169.9}, the rank of 1,000 being 60.9\ignore{13.9}, etc.
With the article view volume of each rank, we estimated the past-30-day views for the adversarial revisions in $D_\mathrm{test}^{(\mathrm{all})}$ to be 55,479,625 (which are 39,300,915 before rank boosting). 
%
%
When estimating the percentage of the drug purchase actions being taken after the impressions of revisions (i.e., view-through rate), we used the web search conversion rate into online drug sales (1\%)~\cite{leontiadis2011measuring}. Here we acknowledge the limitation of this estimation due to the difference between the conversion rates of web search and Wiki search.
Using the parameter setting $R_a$=\$200 based on~\cite{wangdemystifying}, we estimated the revenue of promotional revisions in $D_\mathrm{test}^{(\mathrm{all})}$ of $R$(30-day) =1\%$\times$55 Million$\times$\$200=\$110 Million.



\section{Mitigation and Defense}
\label{sec:defenses}

\noindent\textbf{Coherence detection}.
In MAWSEO, revisions make efforts to maintain the inserted paragraph semantically consistent with the topic and neighboring context of target articles.
However, the language coherence of the joints between the inserted paragraph and its two neighboring paragraphs might still be problematic at the sentence level.
In our study, we observed that, for the two paragraphs around the insertion position, the last sentence of the former paragraph and the first sentence of the latter keep coherence in logic and semantic organization. However, this coherence may be broken up by the inserted paragraph, manifested as that the coherence of the former paragraph's last sentence with the inserted one's first sentence is not as good as that with the latter's first sentence. 

Specifically, we found that, although the inserted paragraph shares a similar topic and semantics with the whole neighboring paragraphs, the language, especially the words used in the sentences, has coherence issues. For example, the last sentence of the former paragraph in one revision sample is ``Fexofenadine (Allegra) may be taken orally to prevent and relieve some of the hives ... has not been evaluated,'' and the first of the inserted paragraph is ``Fexofenadine, sold under the brand name Allegra, is an antihistamine pharmaceutical drug.'' Both sentences introduce the same drug but lack the logic of coherence (i.e., the following sentence does not continue to involve ``hive'', causing a logical interruption).
We also noted that, in the revisions of $D_\mathrm{test}^{(\mathrm{all})}$, the average semantic similarity between the former paragraph's last sentence and the latter's first sentence around the insertion position is 0.60, while that between the former paragraph's last sentence and the inserted promotion one's first sentence is only 0.37, implying that a significant language gap on the insertion joint still exists. 

\begin{figure}[!t]
\centering
\includegraphics[width=8.8cm]{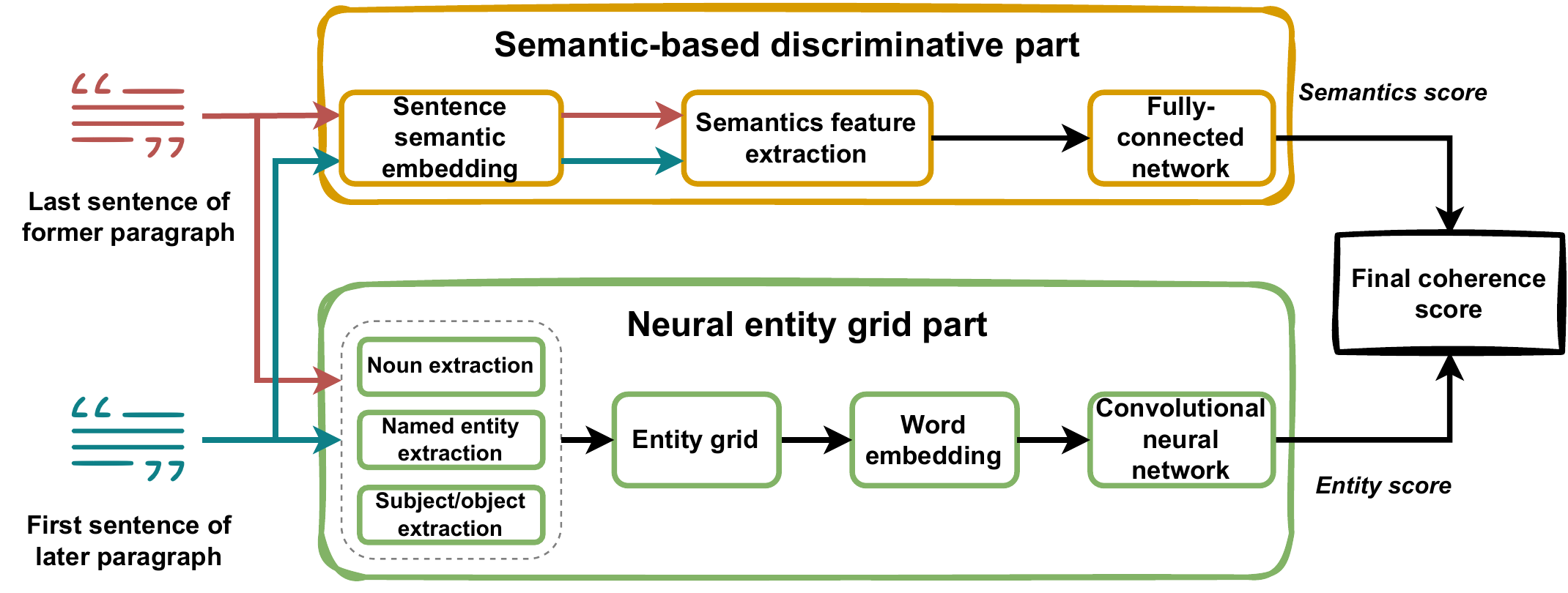}
\caption{Coherence Detection Model.}
\label{fig:defense}
\end{figure}

Many coherence models have been proposed for sentence ordering in which the model is used to distinguish between a coherently ordered sentence list and a random permutation~\cite{nguyen2017neural,joty2018coherence,xu2019cross}. Thus, we can leverage a coherence model to find the revision by distinguishing the disordered sentence list around two revision joints.
Inspired by~\cite{joty2018coherence,xu2019cross}, we design the coherence detection model combining the neural entity grid and the semantic-based discriminative model, depicted in Fig.~\ref{fig:defense}. The neural entity grid captures entity features (e.g., ``hive'' in the first sentence of the above example) and entity transitions (e.g., the disappearance of ``hive'' in the second sentence) along two sentences for entity information (e.g., the first sentence discusses involving ``hive'' while the second does not), which reveals the sentences' logical change in the word level~\cite{cheung2010entity}. 
For semantics information, the semantic-based discriminative model is used for recognizing the semantic relationship between two sentences.
%
After that, our model computes the coherence score based on the combination of the entity and semantics information, by a fully-connected network. 
%
Our model takes a sentence pair and outputs a higher coherence score for a more coherent sentence pair.
We use the pairwise ranking approach to train our model with the objective:
\begin{equation}
\footnotesize
    \mathcal{L}(\theta) = \max\{{0, 1 - f(s_i, s_{i+1}|\theta) + f(s_i, s'|\theta)}\}
\label{equ:pairwiseloss}
\end{equation}
where $f(s_i, s_{i+1}|\theta)$ denotes the transformation of the sentence pair $(s_i, s_{i+1})$ to a coherence score done by the model with parameters $\theta$; $(s_i, s_{i+1})$ is the positive ordered pair and $(s_i, s')$ is the negative disordered pair. In our study, $s'$ is the inserted paragraph, with neighboring paragraphs $s_i$, $s_{i+1}$.

In our experiment, $s$ is the first or last two-sentence chunk in a paragraph to provide sufficient information about the start or end of this paragraph. 
We trained our coherence detection model for 100 epochs with the dataset including 172,000 triplets of $(s_i, s_{i+1}, s')$. $s_i$ and $s_{i+1}$ are extracted from two neighboring paragraphs randomly selected from an article in our local Wiki system, representing the last two-sentence chunk from the former paragraph and the first two-sentence chunk from the latter. $s'$ is the first two-sentence chunk from a paragraph selected randomly from another article. 
For the test, we made revision samples and legitimate samples by extracting the triplets from the revision samples generated by MAWSEO and legitimate Wiki articles, respectively. For revision samples, $s'$ is the first two-sentence chunk of the inserted paragraph, and $s_i$ and $s_{i+1}$ are the last or first two-sentence chunk of two paragraphs around the insertion position. For legitimate samples, $s_i$, $s_{i+1}$, and $s'$ are from an ordered three-paragraph sequence, in which $s_i$ is the last two-sentence chunk of the first paragraph, $s_{i+1}$ is the first two-sentence chunk of the second, and $s'$ is the first two-sentence chunk of the third.

\noindent\textbf{Adversarial training of vandalism detector}.
Improving the vandalism detector's robustness is another way to mitigate MAWSEO revisions. To make the Wiki vandalism detector more robust, we used adversarial training in which the defender includes revision examples generated by MAWSEO into the training data~\cite{kurakin2016adversarial}.
We assume that the defender has sufficient revision examples for training.
In our experiment, we rebuilt the feature-based ORES \texttt{damaging} detection model with the gradient boosting algorithm, the same algorithm used by ORES~\cite{halfaker2020ores}. We trained the rebuilt vandalism detection model with the feature vectors from the vandalism training set in Section~\ref{subsubsec:vandalismData} and the revision results of $D^{(\mathrm{all})}_\mathrm{train}$. We evaluated the model on the revisions of $D^{(\mathrm{t20})}$ and $D^{(\mathrm{all})}_\mathrm{test}$  along with the same amount of legitimate samples randomly selected from the vandalism test set.

\noindent\textbf{Evaluation}.
The evaluation results are listed in Table~\ref{table:defense}, where we can observe that coherence detection is generally more effective in detecting revisions than adversarial training.
The coherence detection successfully identified 87.53\% and 95.09\% of MAWSEO revisions in $D^{(\mathrm{t20})}$ and $D_\mathrm{test}^{(\mathrm{all})}$, respectively. 
%
The detection model performed better on $D_\mathrm{test}^{(\mathrm{all})}$ compared to $D^{(\mathrm{t20})}$.
It is due to the higher semantic consistency in $D^{(\mathrm{t20})}$, as shown in Table~\ref{table:promotionResult}, implying that the revisions in $D^{(\mathrm{t20})}$ are likely to have more fluid paragraph joints.
Also, given that the MAWSEO revisions in $D_\mathrm{test}^{(\mathrm{all})}$ exhibited better evasion ability than those in $D^{(\mathrm{t20})}$ (see Table~\ref{table:promotionResult}), adversarial training was harder to detect the revisions in $D_\mathrm{test}^{(\mathrm{all})}$ than in $D^{(\mathrm{t20})}$, which yielded an accuracy of 50.53\% and 81.72\%, respectively. 
The accuracies of legitimate samples for both approaches were all above 80\% and did not change too much across various search result sets for each defense method.
%
%
Thus, the proposed defense methods can be supplements to the defensive mechanism against Wiki adversarial revisions in the Wiki system.

\begin{table}[!t]
\scriptsize
\centering
\caption{Performance of defense methods}
\label{table:defense}  
\begin{tabular}{l|c|c|c|c}
\hline
Search Result Set & \multicolumn{2}{c|}{Top-20 ($D^{(\mathrm{t20})}$)} & \multicolumn{2}{c}{All ($D^{(\mathrm{all})}_\mathrm{test}$)} \\
\hline
Method & \makecell{Coherence\\Detection} & \makecell{Adversarial\\Training} & \makecell{Coherence\\Detection} & \makecell{Adversarial\\Training} \\ 
\hline
\hline
Accuracy (Revision) & 87.53\% & 81.72\% & 95.09\% & 50.53\% \\
\hline
Accuracy (Legitimate) & 82.38\% & 84.56\% & 81.85\% & 85.17\%  \\
\hline
\end{tabular}
\end{table}

\section{Discussion and Limitations}
\label{sec:discussion}

\noindent\textbf{Discussion}.
As a black-box attack against the real-world Wiki system, our approach not only outperforms baseline approaches (see Section~\ref{subsubsec:baselines}) but also has a higher or similar attack success rate compared with other real-world attacks in natural language processing~\cite{li2018textbugger,he2021model}, whose attack success rates are located between 10\% and 50\%. 

With the rise of Large Language Models like ChatGPT, we utilized GPT-3.5~\cite{gpt35} for generating paragraphs for illicit promotion on Wiki. However, our tests on datasets $D^{(\mathrm{t20})}$ and $D_\mathrm{test}^{(\mathrm{all})}$ showed its poor performance in illicit promotion, with a promotion success rate of only 10.42\% and 8.85\%. The success rates for rank boosting, detection evasion, topic relevancy, and semantic consistency were 28.13\%, 63.31\%, 78.83\%, and 58.85\% for $D^{(\mathrm{t20})}$ and 28.28\%, 84.86\%, 75.33\%, and 52.95\% for $D_\mathrm{test}^{(\mathrm{all})}$. The results indicate that while GPT-3.5 performs well in crafting fluent text relevant to topics and semantics, it still struggles in specialized tasks, like rank boosting and detection evasion, compared with MAWSEO. More details can be found in Appendix~\ref{subsec:prompt}.

\noindent\textbf{Limitations}.
However, our study still has limitations imposed by raw paragraph collecting and topic selection.
First, in our evaluation of MAWSEO, the raw paragraphs were randomly collected from WikiDump. A well-prepared raw paragraph set manually written or designed for one specific task would further strengthen the performance of our proposed approach. It is because MAWSEO can retrieve much more suitable paragraphs from this paragraph set generated based on target queries and target articles.
Second, in our paper, we have discussed the application of MAWSEO on the tasks of illicit online pharmacy promotion and further generalized it to illicit online casino promotion, in which MAWSEO and its defense methods show a good generalization ability (in Appendix~\ref{subsec:gambling}). In the meantime, this approach can be applied to other topics like pornography, politics, law, etc. 
It is also applicable to other ranking-based interactive platforms (e.g., Quora and Reddit) that share similar conditions as outlined in Section~\ref{subsec:threatmodel}.
We leave the applications on these topics and platforms as future work.

\section{Conclusion}
In this paper, we study the robustness of Wiki search engine and state-of-the-art Wiki vandalism detectors against the modern Wiki search poisoning attack, which is equipped with adversarial ranking attack techniques, for illicit online promotion.
In our study, we consider real-world attack objectives of illicit promotion, e.g., rank boosting, vandalism detection evasion, topic relevancy, semantic consistency, and user awareness (but not alarming) of promotional content. 
To this end, this paper presents MAWSEO, a novel blackhat Wiki SEO technique that can effectively and efficiently generate vandalism edits achieving the aforementioned attack objectives. 
%
%
%
Also, we study the potential countermeasures, including sentence-level coherence detection and adversarial training techniques. 
Our discoveries and new techniques have made a step toward a better understanding of Wiki data poisoning for illicit promotion, contributing to a more effective defense against such a threat.

\section*{Acknowledgement}
We sincerely thank our shepherd and the anonymous reviewers for their insightful comments, and Yanxue Jia for the discussions.
This work is supported in part by the NSF CNS-1850725 and 1801432.

\section*{Availability}
The code and dataset are available on request at the project website~\url{https://sites.google.com/view/mawseo}.


\appendices

\section{Design of Baseline Approaches}
\label{app:baselineDesign}
\ignore{
We implemented two language-generation based baseline methods \textit{HotFlip} and \textit{GPT-2} for adversarial ranking attacks\zlccs{~\cite{song2020adversarial,liu2022order}}. 
\zlccs{To make the revisions generated by the baselines satisfy the four attack objectives for the Wiki system environment (see Section~\ref{subsec:threatmodel}) that is aligned with our approach, we alternated the passage retrieval network pipeline with baseline methods to generate adversarial revisions for comparison experiments.}
}
We implemented three language-generation based adversarial ranking attacks \textit{HotFlip}~\cite{ebrahimi2018hotflip,song2020adversarial}, \textit{Collision}~\cite{song2020adversarial}, and \textit{PAT}~\cite{liu2022order} as baseline approaches.

\noindent\textbf{HotFlip}. 
HotFlip, a gradient-based method, crafts adversarial text via token replacement~\cite{ebrahimi2018hotflip}. 
%
%
In HotFlip, adversaries aim to craft text unrelated yet perceived as semantically akin to the target query by the ranking model, boosting the injected article's rank. 
While this work originally assumed that adversaries have full knowledge of the ranking model, to align the attack with our paper's black-box setting, we adjusted the method by allowing adversaries to only access the ranking model's results and the substitute ranker trained on them.
Following prior work~\cite{song2020adversarial}, we used a \ignore{HotFlip-based }greedy method for rank boosting, iteratively replacing words in an initial sequence with repeated words like "this" under constraints from our local substitute ranker. This converts word replacement to an optimization problem that can be solved by minimizing the loss:
\begin{equation}
\footnotesize
    \min{({v_{\widetilde{w}_i}} - {v_{w_i}})^{\top}{\nabla_{v_{\widetilde{w}_i}}}{\mathcal{L}_\mathrm{rank}}}
\label{equ:hotflip}
\end{equation}
where $v_{w_i}$ and $v_{\widetilde{w}_i}$ denote embeddings of the replaced word and the new word at position $i$. Promotional content is injected into this generated paragraph via the incentive injection model to generate a promotion paragraph, which is then inserted between the target article's two paragraphs that are most similar to it in semantics.
%

\noindent\textbf{Collision}. 
Collision~\cite{song2020adversarial}, unlike HotFlip's hard language, uses a gradient-based approach and GPT-2~\cite{radford2019language} to generate natural adversarial text.
Collision shares the same threat model as HotFlip in the context of adversarial ranking attacks.
It optimizes a perturbation $\delta_{i+1}$ added to next-token logits $o_{i+1}$ from GPT-2 at step $i$, with gradients and loss from our substitute ranker. The goal is to find perturbed logits ${\widetilde{o}}{i+1} = o{i+1} + \delta_{i+1}$, whose sampled token $\widetilde{w}{i+1}$---joined with previous tokens $\widetilde{w}_{1:i}$ and target article content $w^c$---minimizes the loss\ignore{ as given by Equation~\ref{equ:subrankloss}.}
:
\begin{equation}
\footnotesize
    \min{\mathcal{L}_\mathrm{rank}(\widetilde{w}_{1:i} \oplus \widetilde{w}_{i+1} \oplus {w^c})}
\label{equ:gpt}
\end{equation}
From the perturbed logits, a token is sampled to create \ignore{the next token's }subsequent logits. 
Using the HotFlip procedure, we create a promotion paragraph and then modify the target article based on the resultant paragraph.


\begin{table}[!t]
\scriptsize
\centering
\caption{Performance comparison of ranking baselines \ignore{on ranking test set}}
\label{table:rankbaseline}  
\begin{tabular}{l|c|c|c}
\hline
Baseline Models & MSE & NDCG@20 & NDCG@200 \\ 
\hline
\hline
\textbf{MV-LSTM} & \textbf{4.59} & \textbf{96.43\%} & \textbf{98.68\%} \\
\hline
BM25 & 74.59 & 95.62\% & 98.11\% \\
\hline
CDSSM & 5.42 & 90.22\% & 96.71\% \\
\hline
DRMM & 12.48 & 70.08\% & 89.90\% \\
\hline
MatchPyramid & 7.75 & 86.38\% & 95.36\% \\
\hline
Duet & 6.60 & 95.60\% & 98.33\% \\
\hline
\end{tabular}
\end{table}

\noindent\textbf{PAT}. We compared our model with a cutting-edge black-box adversarial ranking attack model, the Pairwise Anchor-based Trigger (PAT) generation model~\cite{liu2022order}. 
In PAT's threat model, adversaries aim to strategically inject an optimized adversarial trigger to deliberately disorder rankings. The injected trigger should be contextually consistent and not nonsensical. This attack operates under a black-box setting where adversaries lack access to the ranking model's architecture, training data, and scoring function but can view its ranking results.
\ignore{Like prior work~\cite{song2020adversarial}, }PAT uses gradient-based search to generate a fixed-length trigger text. A substitute ranking model first imitates the target model, and then, with beam search, identifies the trigger while a language model and the next sentence prediction model ensure semantic consistency and fluency. 
Finally, promotional content is injected into the trigger using the incentive injection model and inserted at the beginning of the target article's body as per PAT's trigger injection position~\cite{liu2022order}.

\begin{table*}[!t]
\scriptsize
\centering
\caption{Performance without one discriminator and under uni-constraint on different search result sets}
\label{table:remove1discriminator}  
\label{table:uniconstrainedAttack_main}  
\begin{tabular}{l|l|c|c|c|c|c}
\hline
Search Result Set & \makecell{Removed or Retained\\Discriminator} & \makecell{\% Rank\\Boosting Succ.} & \makecell{\% Evasion\\Succ.} & \makecell{\% Topic\\Relevancy} & \makecell{\% Semantic\\Consistency} & \makecell{\% Promotion\\Succ.} \\ 
\hline
\hline
\multicolumn{7}{l}{\texttt{Without one discriminator (remove a discriminator):}}\\
\hline
\multirow{4}*{Top-20 ($D^{(\mathrm{t20})}$)} & Substitute ranker & \textbf{41.69} & 83.33 & 81.48 & 78.41 & 26.51 \\
\cline{2-7}
~ & Substitute vandalism detector & 51.64 & \textbf{74.86} & 82.60 & 80.41 & 26.30 \\
\cline{2-7}
~ & Topic relevancy discriminator & 49.38 & 85.26 & \textbf{71.00} & 72.31 & 27.43 \\
\cline{2-7}
~ & Semantic consistency discriminator & 46.52 & 85.46 & 72.79 & \textbf{70.44} & 25.76 \\
\hline
\multirow{4}*{All ($D_\mathrm{test}^{(\mathrm{all})}$)} & Substitute ranker & \textbf{59.71} & 92.55 & 60.88 & 52.31 & 25.91 \\
\cline{2-7}
~ & Substitute vandalism detector & 73.17 & \textbf{88.21} & 57.37 & 51.37 & 26.27 \\
\cline{2-7}
~ & Topic relevancy discriminator & 72.93 & 91.74 & \textbf{48.82} & 43.44 & 24.60  \\
\cline{2-7}
~ & Semantic consistency discriminator & 68.94 & 93.02 & 49.82 & \textbf{43.16} & 24.18\\
\hline
\multicolumn{7}{l}{\texttt{Under uni-constraint (retain a discriminator):}}\\
\hline
\multirow{4}*{Top-20 ($D^{(\mathrm{t20})}$)} & Substitute ranker & \textbf{67.35} & 73.80 & 64.90 & 69.39 & 27.58 \\
\cline{2-7}
~ & Substitute vandalism detector & 40.40 & \textbf{89.85} & 64.28 & 61.71 & 21.77 \\
\cline{2-7}
~ & Topic relevancy discriminator & 41.27 & 77.09 & \textbf{81.78} & 76.83 & 25.47 \\
\cline{2-7}
~ & Semantic consistency discriminator & 41.92 & 76.95 & 81.54 & \textbf{79.24} & 25.64 \\
\hline
\multirow{4}*{All ($D_\mathrm{test}^{(\mathrm{all})}$)} & Substitute ranker & \textbf{91.53} & 87.80 & 38.96 & 38.06 & 21.71 \\
\cline{2-7}
~ & Substitute vandalism detector & 56.02 & \textbf{95.14} & 43.30 & 37.11 & 19.91 \\
\cline{2-7}
~ & Topic relevancy discriminator & 60.16 & 88.07 & \textbf{60.74} & 52.20 & 29.06  \\
\cline{2-7}
~ & Semantic consistency discriminator & 60.03 & 88.06 & 60.31 & \textbf{52.79} & 28.56\\
\hline
\end{tabular}
\end{table*}

\section{Dataset and evaluation of sub-models}
\label{app:submodels}

\noindent\textbf{Dataset of incentive injection model}
We utilized the semantic role labeling and NER models from \texttt{AllenNLP}~\cite{allennlp} to label revision entities. Specifically, geolocation adverb modifiers and organization names were identified as the replacement entity, while promotional keywords~\cite{wangdemystifying,kalyanam2017detection} and subjects with article topic keywords were marked as the insertion entity. Two security professionals validated the labeled dataset over seven days, helping annotate 1,030 paragraphs from 390 Wiki articles as the ground truth\ignore{ for the binary-attention based BiLSTM-CRF model}.

\noindent\textbf{Dataset of local ranker}. 
To approximate and train the local substitute ranker, we built a ranking dataset using the search results and their ranking scores achieved by querying our local Wiki system.
%
The queries include 659 drug-related queries described in Section~\ref{subsubsec:dataset} and 200 non-drug medical queries extracted from Wikipedia's popular medical topics~\cite{wikiPopMedicalPages} for generalization.
We randomly selected about 200 articles carrying ranking scores from each query's search results and set these query-article-score triplets as the ranking dataset. The ranking training set includes 105,104 triplets in which the queries consist of the drug-related queries for training (see Section~\ref{subsubsec:dataset}) and the non-drug medical queries. The ranking test set contains 15,682 triplets whose queries are the drug-related queries for testing.

\noindent\textbf{Evaluation of local ranker}. 
We compared our local substitute ranker (MV-LSTM) with several state-of-the-art ranking models, including BM25~\cite{robertson1994some}, CDSSM~\cite{shen2014learning}, DRMM~\cite{guo2016deep}, MatchPyramid~\cite{pang2016text}, and Duet~\cite{mitra2017learning}, in local ranker approximation. 
The pointwise ranking performance was evaluated on the ranking test set using two metrics: Mean Square Error (MSE)~\cite{taylor2008softrank} and Normalized Discounted Cumulative Gain (NDCG)~\cite{jarvelin2017ir}.
MSE was used to assess the regression performance of the pointwise model based on the computed scores, and NDCG measured the ranking performance in which the ranking relied on the scores. Specifically, we evaluated the top 20 and top 200 ranking results by NDCG@20 and NDCG@200, respectively. 
As shown in Table~\ref{table:rankbaseline}, our local substitute ranker outperformed other ranking models.

\ignore{
To further compare the performance in MAWSEO between baselines and our method, we conducted an ablation study by replacing the local substitute ranker with different ranking methods to compare the performance in MAWSEO, especially ranking manipulation and promotion effectiveness. 
The results are provided in Table~\ref{table:rankerApproxAbla}. As we can see, MV-LSTM has the highest rank boosting success rate and promotion success rate on both search result sets.  

\begin{table}[!t]
\scriptsize
\centering
\caption{Performance comparison of baseline ranking models in MAWSEO}
\label{table:rankerApproxAbla}  
\begin{tabular}{l|c|c|c|c}
\hline
Search Results & \multicolumn{2}{c|}{Top-20 ($D^{(\mathrm{t20})}$)} & \multicolumn{2}{c}{All ($D_\mathrm{test}^{(\mathrm{all})}$)} \\
\cline{1-5}
Models & \makecell{\% Rank\\Boost Succ.} & \makecell{\% Promotion\\Succ.} & \makecell{\% Rank\\Boost Succ.} & \makecell{\% Promotion\\Succ.}\\
\hline
\hline
\textbf{MV-LSTM} & \textbf{47.72} & \textbf{28.81} & \textbf{68.75} & \textbf{30.22} \\
\hline
BM25 & 43.25 & 26.83 & 61.35 & 27.38 \\
\hline
CDSSM & 46.60 & 27.99 & 65.32 & 28.06 \\
\hline
DRMM & 46.85 & 28.13 & 66.10 & 27.28 \\
\hline
MatchPyramid & 45.30 & 27.71 & 64.24 & 28.38 \\
\hline
Duet & 47.26 & 28.36 & 66.55 & 28.94 \\
\hline
\end{tabular}
\end{table}
}

\noindent\textbf{Dataset of local vandalism detector}
We built a vandalism dataset to approximate and train the local substitute vandalism detector. 
Specifically, we randomly collected 14,786 articles from the local Wiki system as target articles and randomly collected 8,463 paragraphs from 7,713 non-targeted articles as insertions. 
%
Revisions were created by randomly inserting a collected paragraph into a target article, with labels (damaging or non-damaging) based on ORES \texttt{damaging} model. 
This resulted in 19,531 damaging and 50,469 non-damaging revisions. For balanced training, we randomly reduced non-damaging revisions to 19,531, yielding a balanced vandalism dataset of 39,062 revisions.

\section{Validation Criteria of User Study}
\label{subsec:invalidResponse}

%
Low-quality responses (i.e., invalid responses) are from participants who hastily completed questionnaires without proper attention. We considered such responses invalid.   

To ensure quality, we validated responses based on questionnaire answers, time duration, and completeness.
One criterion for invalid responses (38.9\%) was that participants selected an irrelevant choice in the topic question, indicating a poor-quality answer. Following a common way to control the quality of user-study responses~\cite{Inattentive\ignore{,zhang2022beyond}}, we included an obviously irrelevant choice for each multi-choice topic question. If a participant selected this choice, we view her response as invalid. 
Additionally, we consider responses invalid if participants finished answering the questionnaire within two minutes (27.8\%) or did not complete all the questions of the questionnaires (33.3\%).

\section{Supplementary Details in Ablation Study}
\label{subsec:ablation1Discriminator}
\label{subsec:unicostraintresults}
\label{subsec:alternative}
Table~\ref{table:remove1discriminator} displays the model performance without one discriminator and under uni-constraint.


\section{Study on Illicit Online Casino Promotion}
\label{subsec:gambling}

\begin{table*}[!t]
\scriptsize
\centering
\caption{MAWSEO performance on illicit online casino promotion}
\label{table:GamblePromotionResult}  
\begin{tabular}{l|l|c|c|c|c|c|c}
\hline
Search Result Set & Approach & \makecell{\% Rank\\Boosting Succ.} & \makecell{\% Evasion\\Succ.} & \makecell{\% Topic\\Relevancy} & \makecell{\% Semantic\\Consistency} & \makecell{\% Promotion\\Succ.} & \makecell{Time Cost\\(hrs)}\ignore{\makecell{Ave.\\Time (s)}} \\ 
\hline
\hline
\multirow{4}*{Top-20 (${G}^{(\mathrm{t20})}$)} & \textbf{MAWSEO} & \textbf{44.69} & \textbf{92.83} & \textbf{83.74} & \textbf{75.98} & \textbf{31.60} & \textbf{11.41\ignore{0.16}} \\
\cline{2-8}
 & HotFlip & 26.09 & 68.67 & 2.20 & 0.00 & 0.00 & 28.38\ignore{46.93} \\
\cline{2-8}
 & Collision & 25.26 & 59.90 & 30.45 & 2.39 & 0.18 & 81.29\ignore{134.42} \\
 \cline{2-8}
 & PAT & 25.17 & 81.90 & 2.94 & 0.32 & 0.05 & 144.78\ignore{239.42} \\
\hline
\multirow{4}*{All (${G}_\mathrm{test}^{(\mathrm{all})}$)} & \textbf{MAWSEO} & \textbf{66.13} & \textbf{94.50} & \textbf{69.05} & \textbf{56.94} & \textbf{33.21} & \textbf{11.52\ignore{0.18}} \\
\cline{2-8}
 & HotFlip & 31.90 & 67.16 & 1.44 & 0.12 & 0.00 & 32.92\ignore{48.65}  \\
\cline{2-8}
 & Collision & 32.55 & 58.54 & 23.65 & 1.64 & 0.08 & 97.77\ignore{144.49} \\
 \cline{2-8}
 & PAT & 34.73 & 78.94 & 2.05 & 0.25 & 0.04 & 166.83\ignore{246.54} \\
\hline
\end{tabular}
\end{table*}

\begin{table}[!t]
\scriptsize
\centering
\caption{Performance of defense methods against illicit online casino promotion}
\label{table:gambleDefense}  
\begin{tabular}{l|c|c|c|c}
\hline
Search Result Set & \multicolumn{2}{c|}{Top-20 (${G}^{(\mathrm{t20})}$)} & \multicolumn{2}{c}{All (${G}^{(\mathrm{all})}_\mathrm{test}$)} \\
\hline
Method & \makecell{Coherence\\Detection} & \makecell{Adversarial\\Training} & \makecell{Coherence\\Detection} & \makecell{Adversarial\\Training} \\ 
\hline
\hline
Accuracy (Revision) & 80.20\% & 69.22\% & 84.56\% & 57.84\%  \\
\hline
Accuracy (Legitimate) & 82.04\% & 84.80\% & 80.99\% & 85.14\%  \\
\hline
\end{tabular}
\end{table}

We tested our prototype on illicit online casino promotion to further assess its effectiveness across a different topic.

In the implementation, we extended our local Wiki system with gambling-related articles. 
%
We sourced 71,928 articles from Wikipedia, specifically from the Gambling category (depth of ten) and the Games category (depth of four). 
All other implementation settings of the local Wiki system remain consistent with Section~\ref{subsec:implementation}. 

For illicit online casino promotion, we used 109 online casinos collected from the Kaggle Online Gambling Sites dataset~\cite{kaggleGambling} as target promoted businesses. We also collected 109 popular gambling keywords~\cite{CasinoKeywords,wordstreamCasinoKeywords} as the query terms, with an average length of 2.0. 
Using the same way as Section~\ref{subsubsec:dataset}, we generated two sets from the search results of such queries returned by the local Wiki system: top-20 search result set ${G}^{(\mathrm{t20})}$ and sampled all search result set ${G}^{(\mathrm{all})}$, containing 2,177 and 8,392 query-articles pairs, respectively. We randomly selected 76 (70\%) query terms associated with 5,956 query-article pairs (${G}^{(\mathrm{all})}_\mathrm{train}$) for training. For testing, we used the rest of sampled all search result set (${G}^{(\mathrm{all})}_\mathrm{test}$) and the pairs in the top-20 search result set (${G}^{(\mathrm{t20})}$).
We used ${G}^{(\mathrm{all})}_\mathrm{train}$ to fine-tune the trained binary-DSSM based passage retrieval network with other trained models in MAWSEO (i.e., incentive injection model, local substitute ranker, and local substitute vandalism detector) unchanged. 
Also, we selected 21,015 paragraphs pertaining to gambling subjects as raw paragraphs.

\noindent\textbf{Effectiveness and efficiency on casino promotion}.
%
Table~\ref{table:GamblePromotionResult} reveals that MAWSEO produced 688 (31.60\%) and 809 (33.21\%) adversarial revisions\ignore{ meeting all attack objectives} for ${G}^{(\mathrm{t20})}$ and ${G}^{(\mathrm{all})}_\mathrm{test}$.
It indicates MAWSEO's efficacy in creating adversarial revisions for casino promotions, comparable to its performance in pharmacy promotions (see Section~\ref{subsec:promotionEffective}). 
%
MAWSEO outperforms the baselines (i.e., HotFlip, Collision, and PAT) in both efficacy and efficiency.
Also, A user study on MAWSEO's casino promotion echoed results from \ignore{revisions yielded results similar to }Section~\ref{subsec:userstudy}. Further details are available on the \href{https://sites.google.com/view/mawseo}{project website}.
Thus, MAWSEO has a generalization ability to craft adversarial revisions for other illicit online promotions.

\noindent\textbf{Defense performance}.
We assessed two defense models proposed and trained in Section~\ref{sec:defenses} against MAWSEO revisions for casino promotion.
As per Table~\ref{table:gambleDefense}, they effectively counteract MAWSEO revisions, mirroring their efficacy against pharmacy promotion in Table~\ref{table:defense}.
%
%
It also indicates their good generalization in defending MAWSEO revisions for other illicit online promotions.

\section{Prompt of Experiment on GPT-3.5}
\label{subsec:prompt}
\noindent\textbf{Prompt}: {``Please write a paragraph carrying [Promotional content] which can be inserted into the article [Article text]. 
This paragraph should be in Wiki style, and can boost the article's rank in Wiki search engine when searching with query [Target query], evade the vandalism detector, maintain topic relevancy and semantic consistency.''}

\end{document}